\documentclass[conference]{IEEEtran}

\let\labelindent\relax %
\usepackage{enumitem}
\usepackage{xspace}
\usepackage{pgfplots}
\usepackage{pgfplotstable}
\usepackage{tikz}
\usepackage{graphicx,amsmath}
\usepackage{booktabs}
\usepackage{xcolor}
\usepackage{hyperref}
\usepackage{wrapfig}
\usepackage{amsfonts}
\usepackage{comment}
\usepackage{makecell}
\usepackage{multirow}
\usepackage{pifont}
\usepackage{float}
\usepackage{tcolorbox}
\usepackage{subcaption}
\usepackage{csvsimple}
\usepackage{booktabs}
\usepackage{threeparttable}
\usepackage{multicol,calc}
\usepackage{mdframed}
\usepackage[capitalise]{cleveref}
\usepackage{flushend}
\usepackage{siunitx}
\usepackage{url}

\graphicspath{{figures/}}

\newcommand\para[1]{\vspace{2pt} \noindent \textbf{#1.}}

\makeatletter
\newcommand\breaktuple[1]{%
  \@tempcnta=0
  (
  \@for\@ii:=#1\do{%
    \@insertbreakingcomma
    \@ii
  }%
  )
}
\def\@insertbreakingcomma{%
  \ifnum \@tempcnta = 0 \else\,,\ \linebreak[1] \fi
  \advance\@tempcnta\@ne
}
\makeatother

\newcommand*{\priority}[1]{\begin{tikzpicture}[scale=0.09]%
  \draw (0,0) circle (1);
  \fill[fill opacity=1,fill=black] (0,0) -- (90:{#1>0?1:0}) arc (90:90-#1*3.6:1) -- cycle;
  \end{tikzpicture}}

\def\name{\textsf{PayOff}}

\newcommand\property[1]{\textsf{#1}}
\newcommand{\tablehead}[2]{\multicolumn{1}{@{}>{\centering\arraybackslash}p{#1}@{}}{\textbf{#2}}}

\begin{document}

\title{\name{}: A Regulated Central Bank Digital Currency with Private Offline Payments}

\author{\IEEEauthorblockN{Carolin Beer\IEEEauthorrefmark{1}, Sheila Zingg\IEEEauthorrefmark{1}, Kari Kostiainen\IEEEauthorrefmark{1}, Karl W\"ust\IEEEauthorrefmark{2}, Vedran Capkun\IEEEauthorrefmark{3}, Srdjan Capkun\IEEEauthorrefmark{1}}
\IEEEauthorblockA{\IEEEauthorrefmark{1}ETH Zurich
    \\\{carolin.beer, sheila.zingg, kari.kostiainen, srdjan.capkun\}@inf.ethz.ch}
\IEEEauthorblockA{\IEEEauthorrefmark{2}Mysten Labs
    \\karl@mystenlabs.com}
\IEEEauthorblockA{\IEEEauthorrefmark{3}HEC Paris
     \\capkun@hec.fr}
}

\maketitle

\begin{abstract}
The European Central Bank is preparing for the potential issuance of a central bank digital currency (CBDC), called the digital euro. A recent regulatory proposal by the European Commission defines several requirements for the digital euro, such as support for both online and offline payments. Offline payments are expected to enable cash-like privacy, local payment settlement, and the enforcement of holding limits. While other central banks have expressed similar desired functionality, achieving such offline payments poses a novel technical challenge. We observe that none of the existing research solutions, including offline E-cash schemes, are fully compliant. Proposed solutions based on secure elements offer no guarantees in case of compromise and can therefore lead to significant payment fraud. 

The main contribution of this paper is \name{}, a novel CBDC design motivated by the digital euro regulation, which focuses on offline payments. We analyze the security implications of local payment settlement and identify new security objectives. \name{} protects user privacy, supports complex regulations such as holding limits, and implements safeguards to increase robustness against secure element failure. Our analysis shows that \name{} provides strong privacy and identifies residual leakages that may arise in real-world deployments. Our evaluation shows that offline payments can be fast and that the central bank can handle high payment loads with moderate computing resources. However, the main limitation of \name{} is that offline payment messages and storage requirements grow in the number of payments that the sender makes or receives without going online in between. 
\end{abstract}

\begin{table*}
\begin{center}
\footnotesize
\newcommand*\rot[1]{\hbox to1em{\hss\rotatebox[origin=br]{-35}{#1}}}
\newcommand*\feature[1]{\ifcase#1 \priority{0}\or\priority{50}\or\priority{100}\or -\fi}
\newcommand*\fa[4]{\feature#1&\feature#2&#3&\feature#4}
\newcommand*\fb[4]{\feature#1&\feature#2&\feature#3&\feature#4}
\newcommand*\fc[3]{\feature#1&\feature#2&\feature#3}
\newcommand*\fd[5]{\feature#1&\feature#2&\feature#3&#4&\feature#5}
\makeatletter
\newcommand*\ex[7]{#1\tnote{#2}&#3&\fa#4&\fb#5&\fc#6&\fd#7}
\makeatother
\begin{threeparttable}
\begin{tabular}{@{}p{0.16\textwidth}p{0.05\textwidth} c c c c !{\kern1.3em} c c c c !{\kern1.3em} c c c !{\kern1.3em} c c c c c@{}}
& \rot{Fund representation}
& \rot{Offline payments}
 & \rot{Light client support}
 & \rot{Scalability}
 & \rot{Transferability}
 & \rot{Held money privacy}
 & \rot{Payment confidentiality}
 & \rot{Unlinkability}
 & \rot{Offline session unlinkability}
 & \rot{Holding limits}
 & \rot{Receiving limits}
 & \rot{Sending limits}
 & \rot{Counterfeit detection}
 & \rot{Counterfeit creators identifiable}
 & \rot{Counterfeit colluders identifiable}
 & \rot{Risk carrier for fraud}
 & \rot{Robust to secure element failure}\vspace{-8pt}\\
\midrule
\ex{Cash}{} {Coin} {23{High}2} {2222} {000} {200{Recipient}3}\\
\midrule
\ex{Online E-cash~\cite{chaum1983blind}}{} {Coin} {02{High}3} {1113} {021} {333{-}3}\\
\ex{Tourbillon~\cite{tourbillion}}{} {Coin} {02{High}3} {1113} {021} {333{-}3}\\
\ex{Offline E-cash~\cite{chaum_1990_untraceable}}{} {Coin} {22{High}0} {1111} {001} {220{Recipient}2}\\
\ex{Transferable E-cash~\cite{canard2008anonymity}}{} {Coin} {22{Medium}2} {2111} {000} {220{Recipient}2}\\
\ex{Zerocash~\cite{sasson2014zerocash}}{} {UTXO} {01{Medium}3} {2223} {000} {333{-}3}\\
\ex{Regulated Zcash~\cite{garman2017accountable}} {*} {UTXO} {01{Medium}3} {2223} {002} {333{-}3}\\
\ex{Auditable Tokens~\cite{androulaki2020privacy}} {} {UTXO} {01{High}3} {2223} {000} {333{-}3}\\
\ex{UTT~\cite{tomescu2022utt}}{} {UTXO} {01{High}3} {2223} {002} {333{-}3}\\
\ex{Platypus~\cite{wust_2022_platypus}} {}  {Account} {02{High}3} {2223} {222} {333{-}3}\\
\ex{PEReDi~\cite{kiayias_2022_peredi}} {} {Account} {02{High}3} {2223} {222} {333{-}3}\\
\ex{OPS w/o TEE~\cite{christodorescu_2020_twotier}} {} {Account} {22{High}0} {2000} {002} {333{-}0}\\
\ex{OPS w/ TEE~\cite{christodorescu_2020_twotier}} {} {Account} {22{High}2} {2222} {002} {333{-}0}\\
\midrule
\ex{Our solution: \name} {} {Account} {22{Medium}2} {2212} {222} {222{Central Bank}2}\\
\bottomrule
\end{tabular}
\begin{tablenotes}
\item \hfil$\feature2=\text{Yes}$; 
$\feature1=\text{Partial}$; 
$\feature0=\text{No}$;
$\text{\feature3}=\text{Not applicable}$\\
\textsuperscript{*}Coin tracing can make transactions involving specific coins linkable or non-confidential.
\end{tablenotes}    
\caption{Comparison of payment and CBDC systems across dimensions derived from the digital euro regulation~\cite{euro-regulation}.}
\vspace{-10pt}
\label{tab:protocols}
\end{threeparttable}
\end{center}
\end{table*}

\section{Introduction}

The idea of a central bank digital currency (CBDC) has gained increasing attention. Over 100 central banks have ongoing investigations and pilot projects on the topic~\cite{atlantic-council-tracker}. Notably, the European Central Bank (ECB) has communicated its plans to introduce a digital euro~\cite{digital-euro-web} as \emph{retail CBDC}, i.e., accessible to the general public, and the European Commission has recently released a proposal~\cite{euro-regulation} on its regulation. 

We analyzed the regulation proposal, which considers two types of payments. The first is online payments that are settled online by the central bank. Online payments offer similar privacy to bank accounts: Personally identifiable information (PII) for payments is visible to the Payment Service Provider (PSP) of the user, but hidden from the central bank. In such a setting, known anonymous payment techniques could address main privacy concerns, but the design of a highly scalable online CBDC system still requires careful consideration.

The second type is offline payments, where both the sender and recipient can be offline. The regulation mentions security requirements for offline payments, such as cash-like privacy (even towards PSPs), and regulation goals, such as enforcement of holding limits. Furthermore, offline payments should be settled locally on end-user devices, which we call \emph{offline settlement}. As the regulation places no restriction on the use of funds received offline, offline payments should be \emph{transferable}. We note that some users might prefer offline payments, even if they have network connectivity, due to better privacy~\cite{bundesbank-survey}.

In this paper, we focus on regulation-compliant offline retail CBDC payments. Realizing secure offline payments has been noted as a core challenge by recent central bank documents~\cite{bank-of-finland-doc,bank-of-canada-doc}. We also consider this an open problem for the following reasons.

Firstly, simple solutions based on secure elements are insecure. The regulation proposal hints at an approach where offline payments fully rely on secure elements. While secure elements can be a useful building block for a CBDC system, assuming that no secure element gets compromised would be naive. Without further protections, even a single compromised device could print counterfeit money without restrictions.

Secondly, we observe that the offline settlement and transferability requirements, have significant security implications. Offline payment recipients in such a system cannot distinguish legitimate payments from counterfeit and the central bank takes over the risk of fraud. This will likely lead to attacks, such as, an adversary breaking one secure element and sending counterfeit money to a colluding device. We argue that novel security guarantees are needed, such as de-anonymizing counterfeit recipients.

Thirdly, we note that none of the existing research solutions were designed for a similar offline setting, or fulfill all above requirements. UTXO-based anonymous payment systems like Zcash~\cite{sasson2014zerocash}, and its accountable variants~\cite{garman2017accountable}, only support online payments. Similarly, account-based CBDC systems either focus on online settings (Platypus~\cite{wust_2022_platypus} and Peredi~\cite{kiayias_2022_peredi}), or rely on secure elements~\cite{christodorescu_2020_twotier}. Coin-based E-cash solutions have been designed for offline use~\cite{baldimtsi2015anonymous,bauer_2021_transferable,camenisch_2005_compact}, but these solutions do not provide cash-like privacy, are not designed for offline settlement, and do not implement regulation goals like holding limits. \Cref{tab:protocols} provides a detailed comparison.

Our main contribution is \name{}, a novel CBDC solution specifically designed for offline payments and motivated by the digital euro regulation~\cite{euro-regulation}. \name{} is the first solution that (1) focuses on offline payments with offline settlement, (2) provides strong privacy protection for users, (3) implements an extensive set of regulation guarantees, and (4) is robust against secure element failure. We take account-based CBDCs~\cite{wust_2022_platypus,kiayias_2022_peredi} as the starting point of our solution
because they have been shown to be a good fit for deployments that require expressive regulation and strong privacy.

\name{} achieves strong privacy protections: Payment amounts and user identities are confidential, payments are unlinkable across offline sessions and are cryptographically unlinkable within sessions. Residual leakage (e.g., from ledger access patterns) is limited and comparable to other anonymous payment systems. %

Our solution also provides extensive support for regulation. Specifically, we support holding limits, which are present in different regulation proposals~\cite{BoE-technology,euro-regulation}. Furthermore, \name{} implements fraud and double spend detection for offline payments. Although mentioned in recent ECB documents~\cite{ecb-slides, europeancentralbank_2024_progress}, this mechanism has been overlooked in regulations due to the planned reliance of offline systems on secure elements. 
Finally, 
we implement new protection mechanisms like the de-anonymization of counterfeit recipients.  

We evaluate the performance of \name{} through a prototype implementation. The client-side computation for offline payments is fast (e.g., 0.3s). With modest computing resources (e.g., 35 cores), the central bank can process up to 5000 payments per second. The communication requirements grow the longer users stay offline. For offline sessions of a few days, message sizes for offline payments are in the order of a few kilobytes, which still allows fast payments assuming NFC as communication channel. 
Very long offline periods, such as one month, need a higher throughput than NFC can facilitate but could be achieved by leveraging other channels like Bluetooth. %

\section{Background \& related work}

In this section, we provide background on digital currencies and recent CBDC regulations.

\subsection{Privacy-Preserving Digital Payment Systems}
\label{sec:background-currencies}

\para{Coin-based systems} Chaum~\cite{chaum1983blind} introduced E-cash, a digital currency, where funds are represented as \emph{coins}. The classic E-cash design assumes that the payment recipient is online. Follow-up work~\cite{chaum_1990_untraceable,camenisch_2005_compact,canard_2007_divisible}, has adapted E-cash for \emph{offline payments}. As preventing double spending in an offline setting is considered impossible, the main security goal of such works is to detect double spends and \emph{de-anonymize} the dishonest party. Offline E-cash is not \emph{transferable} meaning that the recipient needs to go online and deposit the coins before they can use them. Transferable E-cash~\cite{okamoto_1990_disposable, canard2008anonymity,baldimtsi2015anonymous, bauer_2021_transferable} overcomes this and allows payment recipients to use received coins without needing to go online. 

Coin-based systems have three main drawbacks. 1) While the payment sender can remain anonymous, the recipient cannot. The bank learns the identity of the recipient and the payment value when a coin is deposited. 2) Complex regulations, such as holding limits, cannot be enforced easily. 3) The size of a coin increases in the number of offline payments, potentially making payments slow during long offline periods.

\para{UTXO-based systems} In UTXO-based systems, funds are represented as Unspent Transaction Outputs (UTXOs) and published in a distributed ledger. Zerocash~\cite{sasson2014zerocash} introduced payments with strong privacy by storing cryptographic commitments to outputs instead of plaintext UTXOs on the ledger. The validators, who ensure that each transaction is correctly executed, neither learn the identities of the payment sender and recipient nor the payment value. In addition, validators cannot correlate payments making them \emph{unlinkable}. Zerocash does not consider the enforcement of regulations. Follow-up work, such as~\cite{garman2017accountable, androulaki2020privacy, tomescu2022utt} provide UTXO-based payment systems that can enforce some policies, such as, sending limits, auditability, and de-anonymization for large payment volumes.

UTXO-based systems have two main drawbacks: First, since ledger access is needed, to the best of our knowledge, no existing UTXO-based system supports offline payments. Second, complex regulation, such as holding limits, cannot be easily enforced. 

\para{Account-based systems} Account-based systems like Platypus~\cite{wust_2022_platypus} and Peredi~\cite{kiayias_2022_peredi} represent funds as a balance that is updated with each transaction and balance updates are proven correct in zero-knowledge. This approach provides strong privacy and unlinkability: Neither the identities of the payment sender and recipient nor the transaction values are leaked to the bank. Expressive regulation, such as holding limits, is easier to implement since all user funds are captured by their account balance. OPS~\cite{christodorescu_2020_twotier} is an account-based offline payment system whose security relies on secure elements.

\subsection{CBDC Motivation and Regulation}

More than 100 countries have plans for the introduction of a CBDC~\cite{atlantic-council-tracker}. Some countries, such as China, are already running pilot projects~\cite{MIT-review-US}. The United States and England have published working papers~\cite{BoE-technology} and discussion papers~\cite{fed-paper}. Also, the European Central Bank has published its plans to introduce a digital euro~\cite{digital-euro-web}.

In many places, cash has already largely been replaced by digital payments, such as credit card and mobile payments. However, digital payments do not reach the unbanked population who will face issues if cash becomes obsolete. Furthermore, the cost and risk of relying on large companies for payment services is high. Past incidents of Mastercard and Visa outages have shown the fragility of the current digital payment landscape~\cite{arnold_2018_mastercard}. The deployment of retail CBDC is expected to address these concerns by allowing users to make digital payments with central bank money.

\begin{figure}[t]
    \centering
    \includegraphics[clip, trim=5cm 1.5cm 5cm 2cm, width=0.8\linewidth]{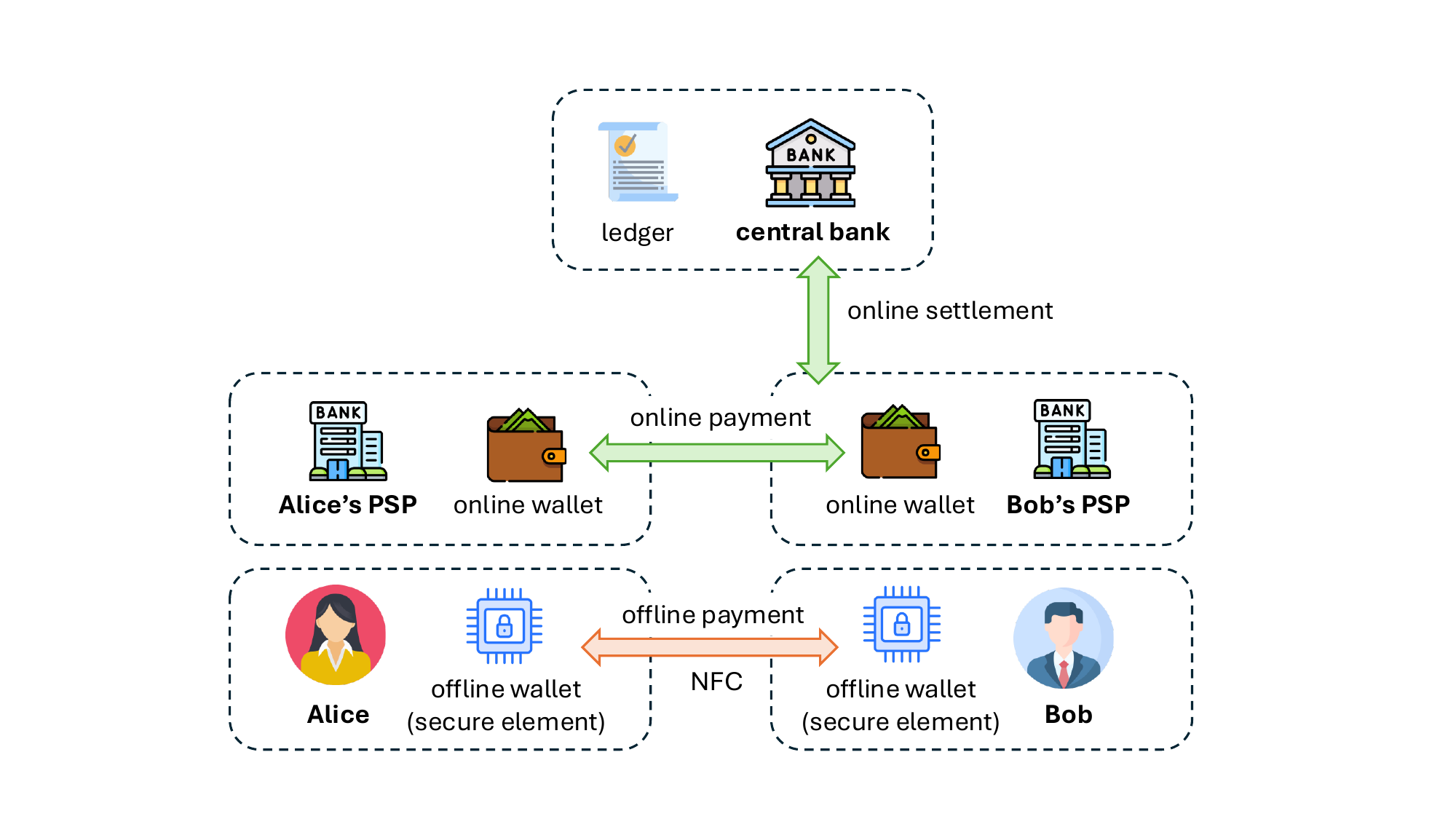}
    \caption{Overview of the digital euro regulation~\cite{euro-regulation}.}
    \vspace{-10pt}
    \label{fig:regulation}
\end{figure}

A recent regulation proposal by the European Commission~\cite{euro-regulation} defines requirements for the digital euro. We summarize the main points of this regulation, depicted in \cref{fig:regulation}.

\para{Entities} \emph{Users} can hold digital euros and send and receive payments. The \emph{central bank} issues money and enforces regulations. The users generally do not interact with the central bank directly, but instead interact with \emph{Payment Service Providers} (PSPs), such as commercial banks.

\para{Payment and wallet types} The regulation distinguishes between \emph{online} and \emph{offline payments}. For each payment type, the user opens a \emph{wallet} with a PSP. Online payments require approval by the sender and recipient PSPs and are settled by the central bank who records the payment on a \emph{ledger}. For offline payments, wallets are hosted on a \emph{secure element} such as a smart card or a Trusted Execution Environment (TEE)~\cite{tee-sok}. Offline payments are restricted to payments in physical proximity and are settled locally on the offline wallets.  

\para{Privacy and security requirements} 
For online payments, the regulation aims for privacy that is similar to existing bank accounts. While PSPs can see the payment details, the central bank should not learn any Personal Identifiable Information (PII). In contrast, offline payments should offer privacy that is comparable to cash. 
A notable regulatory requirement is that users can only hold a limited amount of Digital Euros to avoid destabilizing the financial system.

\section{Problem Statement}

Our goal is to design a solution that supports the requirements for offline payments in regulated CBDCs.

\subsection{Offline Payment Functional Requirements} 

The digital euro regulation states that offline payments should be ``settled locally'' with secure elements, which we interpret as the central bank carrying the risk of potential payment fraud. From this, we identify the following functional requirements (F) for offline payments: 

\para{F1: Offline settlement} If a user correctly follows the protocol and accepts a payment offline, then they are guaranteed to keep the received money even if it was counterfeit.

\para{F2: Transferability} Users who receive payments offline can use the received funds to make further offline payments without connecting to the central bank first.

\subsection{Reliance on Secure Elements} 
A natural solution to offline settlement and transferability is to deploy secure elements to prevent counterfeit creation and other types of fraud, enforce regulations, and preserve user privacy. However, such a solution relies on the assumption that secure elements cannot be compromised, as a compromised device would allow an adversary to create an infinite amount of counterfeit money undetectedly.
Many attacks against secure elements~\cite{smart-card-attacks} and TEEs~\cite{van2022sok, cerdeira2020sok} have been demonstrated. Thus, it is naive to assume that secure hardware for CBDC would not be compromised given the financial benefits. Furthermore, rollback attacks~\cite{matetic2017rote} are often not covered by TEE security guarantees but sufficient to create counterfeits. Lastly, restricting payments to proximity connections is unlikely to limit fraud, as payments could be \emph{relayed} over long distances.

\subsection{Research Question}

Our goal is to design a regulation-compliant offline payment solution that has built-in protections to limit fraud caused by compromised secure elements. 
The prior research that comes closest to our scenario is transferable E-cash~\cite{baldimtsi2015anonymous, canard2008anonymity} where the bank can detect counterfeit money and identify counterfeit money creators when a coin is deposited. However, transferable E-cash still has major limitations. Firstly, E-cash only protects payment sender anonymity, which provides less privacy than cash. Secondly, E-cash does not enforce regulatory rules like holding limits. Lastly, transferable E-cash relies on online settlement, whereas the regulation assumes offline settlement which shifts the risk from payment recipients to the central bank. 
To the best of our knowledge, none of the known solutions fulfills our goals (see \cref{tab:protocols}).

\section{Security Goals}
\label{sec:properties-and-requirements}

In this section, we define more detailed security goals for regulated offline CBDC payments. We consider security goals mentioned in the Digital Euro regulation and identify new goals that we consider necessary to limit payment fraud and hold malicious parties accountable. 

\subsection{Privacy Goals} 
\label{subsec:privacy-security}

The regulation proposal states that offline payments should provide cash-like privacy guarantees. %
We consider the following privacy goals (P) relevant for an offline CBDC:

\para{\property{P1}: Payment and balance confidentiality} Only the user should know how much money they hold. For payments, the value should be only known to the sender and the recipient, and the identities of sender and recipient should remain hidden from third parties.

Users may send or receive more than one payment while offline. Once users go back online, they \textit{reconnect} to the central bank by uploading the payment details for the corresponding payments.

\para{\property{P2a}: Payment unlinkability (towards the central bank)} When a user reconnects to the central bank, the central bank should not be able to link offline payments if the user followed the protocol and did not receive any double spends. 

\para{\property{P2b}: Session unlinkability (towards the central bank)}
A weaker but still useful privacy notion is the unlinkability of offline payments across offline sessions. This means that payments that a user made before and after reconnecting should be unlinkable if the user followed the protocol and did not receive any double spends. 

\para{\property{P3}: Session unlinkability (towards other users)} 
Other users should not be able to link offline payments that a user made before reconnecting, to those that the user made after reconnecting.

We consider the privacy goal \property{P2b} new to the literature of anonymous offline payment systems.

\subsection{Security Goals}

The regulation does not explicitly mention any payment security goals (S), but we consider the following two necessary:	 

\para{\property{S1}: Payment security} Only the user that owns the funds is able to use them for payments. We note that in a CBDC system the central bank issues new money and can therefore make payments without limitations. 

\para{\property{S2}: Payment integrity} Users who have \emph{not} compromised their secure elements should not be able to use more funds than they own. No user should receive more funds than what has been sent to them.

\subsection{Accountability Goals} 
\label{subsec:accountability}

We define four accountability goals (A). The regulation proposal mentions holding limits as one goal. We observe that the definition of ``holding'' money is subtle, since users can receive offline payments but delay the associated message delivery to their offline wallet, and therefore delay the received money from being added to the user balance. This is possible without compromising the secure element. 
	
\para{\property{A1}: Holding limit integrity} We consider a user to \emph{hold} funds when the user can send the funds to honest recipients without relying on the availability of other parties. Users should not be able to hold more funds than their predefined holding limit.

\begin{figure}[t]
    \centering
    \includegraphics[clip, trim = 5cm 1.75cm 6.5cm 2.5cm, width=\linewidth]{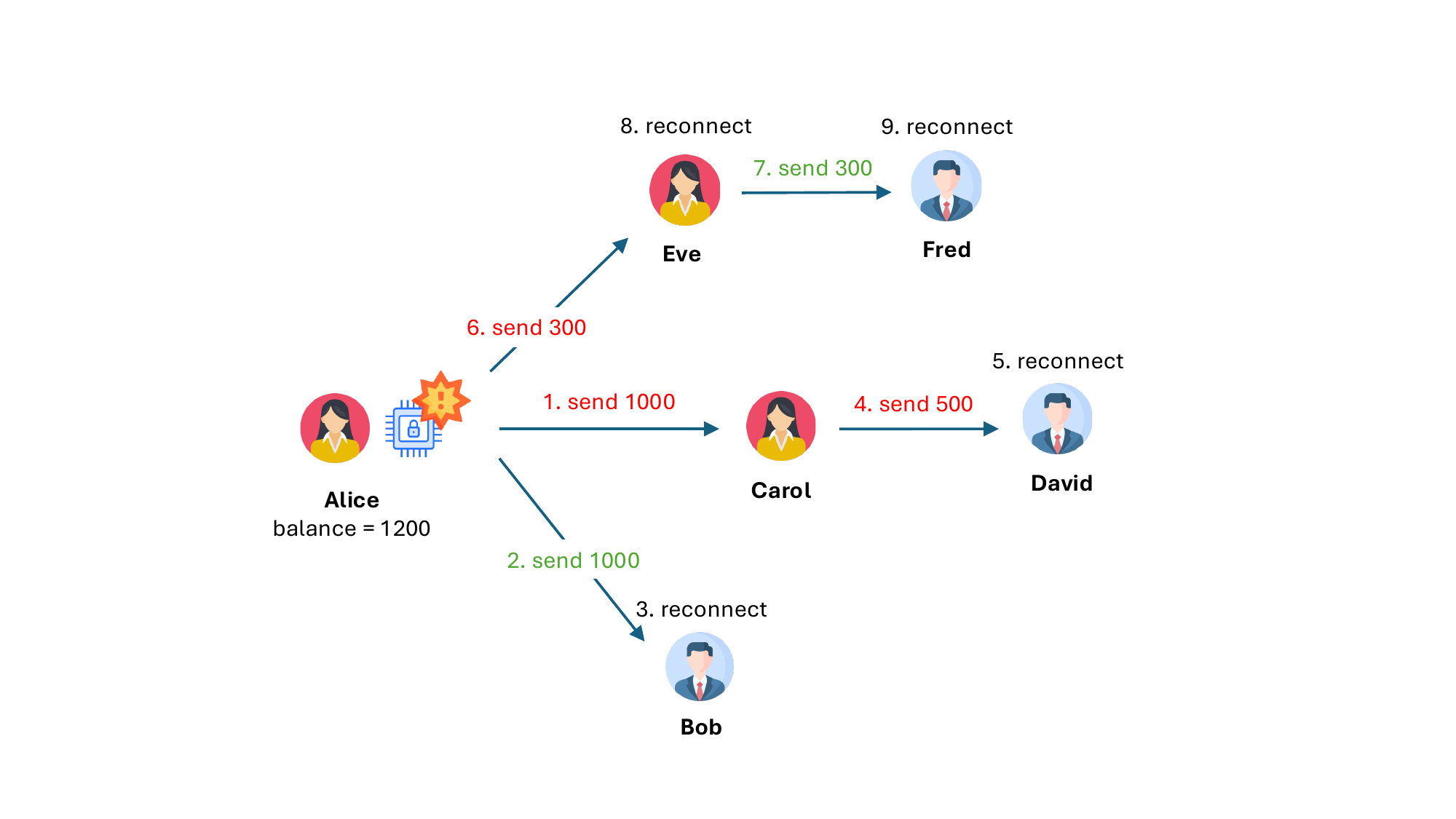}
    \caption{Example payment scenario where Alice double spends.}
    \vspace{-10pt}
    \label{fig:forwarding}
\end{figure}

We now explain what constitutes a \emph{counterfeit} in a payment system that combines offline settlement and double-spend detection, how counterfeits propagate, and which assets need to be traced to detect malicious actions effectively. 

\Cref{fig:forwarding} illustrates an example scenario where Alice has compromised her secure element, and has an initial balance of 1200 in her offline wallet. Alice double spends by creating a digital copy of her funds: she first sends 1000 to Carol, and then sends the same 1000 to Bob. Bob reconnects to the central bank soon after. In such cases, we consider Bob's funds \emph{legitimate} because the central bank has not yet seen the conflicting second payment to Carol when he reconnects. The payment that Carol received is considered \emph{counterfeit}, although she was paid first. Carol can keep the received money, but her assets must be traced.

Next, in our example, Carol sends 500 to David without reconnecting, and after that, David reconnects. David's funds are considered counterfeit because he received a payment from Carol. This example shows how counterfeit propagates.
Finally, Alice creates a third double spend payment to Eve, who pays Fred afterwards. Eve then reconnects and de-anonymizes herself to the central bank. After this point, tracing is no longer needed and Eve's assets are no longer considered counterfeit (to prevent endless propagation). When Fred reconnects, his money is considered legitimate.

In all of these cases, recipients are entitled to keep the funds they received even if they were counterfeit due to the offline settlement requirement. This is different from E-cash and means that double spends increase the total money supply. 

Similarly to previous E-cash literature~\cite{baldimtsi2015anonymous, canard2008anonymity}, our goal is to identify the counterfeit creator when recipients reconnect. For example, in the above example, Alice should be identified once Bob and David reconnect. We define this as follows: 

\para{\property{A2}: Counterfeit creator identification} If a user (who has compromised one or more secure elements) creates counterfeit money, the central bank should (i) detect the fraudulent payment event and (ii) identify the payment creator when at least two recipients reconnect.

Because offline settlement removes the risk that payment recipients lose money due to fraud, we consider collusion between users a relevant threat to the CBDC system. For instance, Alice, Bob, Carol, and Eve could collude to increase their combined assets. We thus argue that the recipients of counterfeit money should also be de-anonymized.%

\para{\property{A3}: Counterfeit recipient de-anonymization} 
If counterfeit money was created, at least one recipient must de-anonymize themselves, reveal the sequence of payments through which they received the money, and disclose the received amount.

Users with compromised secure elements could \emph{omit} certain payments when they reconnect, for instance, uploading all the legitimate payments but leaving out the fraudulent ones.
It is not possible to detect such misbehavior in all cases. For example, if a malicious user omits a payment in which they receive counterfeit funds that they never use, we see no way to detect such an omission. However, this is not beneficial for the user, and our goal is to detect practically relevant omissions. 

\para{\property{A4}: Incomplete synchronization detection} Upon synchronization, if a malicious user omits payments and continues to use their offline wallet with honest users, the central bank will detect this and can de-anonymize the malicious user.

To the best of our knowledge, accountability goals similar to \property{A3} and \property{A4} have not been previously considered.

\section{Solution Overview}
\label{sec:overview}

In this section, we describe our solution gradually, starting from a simplified design that allows only a single offline payment, and then extending it to a complete solution. %

\subsection{Assumptions}
\label{subsec:design-assumptions}

Users trust the central bank to protect the integrity of the money supply, but do not trust it to protect their privacy.  We assume that users can communicate with the central bank anonymously, which is a common assumption in anonymous payment systems~\cite{wust_2022_platypus,kiayias_2022_peredi, sasson2014zerocash}. In practice, anonymous communication could be facilitated by PSPs, mix networks, or similar means. %
Users can drop, delay, or modify messages. However, only users who compromise their secure element can deviate arbitrarily from the protocol. 
We assume that all cryptographic primitives are secure with respect to their security definitions, and the chosen Zero Knowledge Proof (ZKP) system provides weak simulation-extractability, and in case of a trusted setup, subversion-resistance\footnote{This means that the zero knowledge property of the ZKP system holds even if the central bank is corrupted at setup time}. 

\subsection{Representation of Funds}
We follow the account model~\cite{wust_2022_platypus, kiayias_2022_peredi, gross_2021_designing} that was recently proposed for anonymous online payment systems. In this model, the account state is represented by the openings (e.g., the user's balance) of a \emph{state commitment}. Each state is accompanied by a pseudorandom \textit{serial number} that is used to detect the reuse of old states, i.e., double-spending. A payment involves a state transition by the sender and by the recipient. For this, both users update their balance, which results in a new state commitment, and prove in zero-knowledge that their transition is correct, given a commitment to the payment value and some previous state with a valid signature.  
In account-based online systems~\cite{wust_2022_platypus, kiayias_2022_peredi, gross_2021_designing}, both state updates are submitted directly to the central bank. The central bank checks both state updates by verifying the ZKPs and by ensuring that it has not previously seen a state with the same serial number. If all checks pass, the central bank returns a signature over the new state commitments to the users. %

The account model has certain advantages. Firstly, this approach can provide strong privacy (\property{P1-P3}). %
Secondly, a system where all funds are captured by one account balance enables a simple and efficient implementation of regulatory rules such as holding limits (\property{A1}). %
Thirdly, %
since user states evolve as a \emph{sequence} of state transitions, it is possible to reason about \emph{all} payments that of a user between two points in time. This facilitates the detection of omitted payments (\property{A4}).

\subsection{Starting Point: Single Payment}
\label{subsec:single-payment}

\begin{figure}[t]
    \centering
    \includegraphics[clip, trim=5.5cm 1.7cm 5.5cm 1.7cm, width=\linewidth]{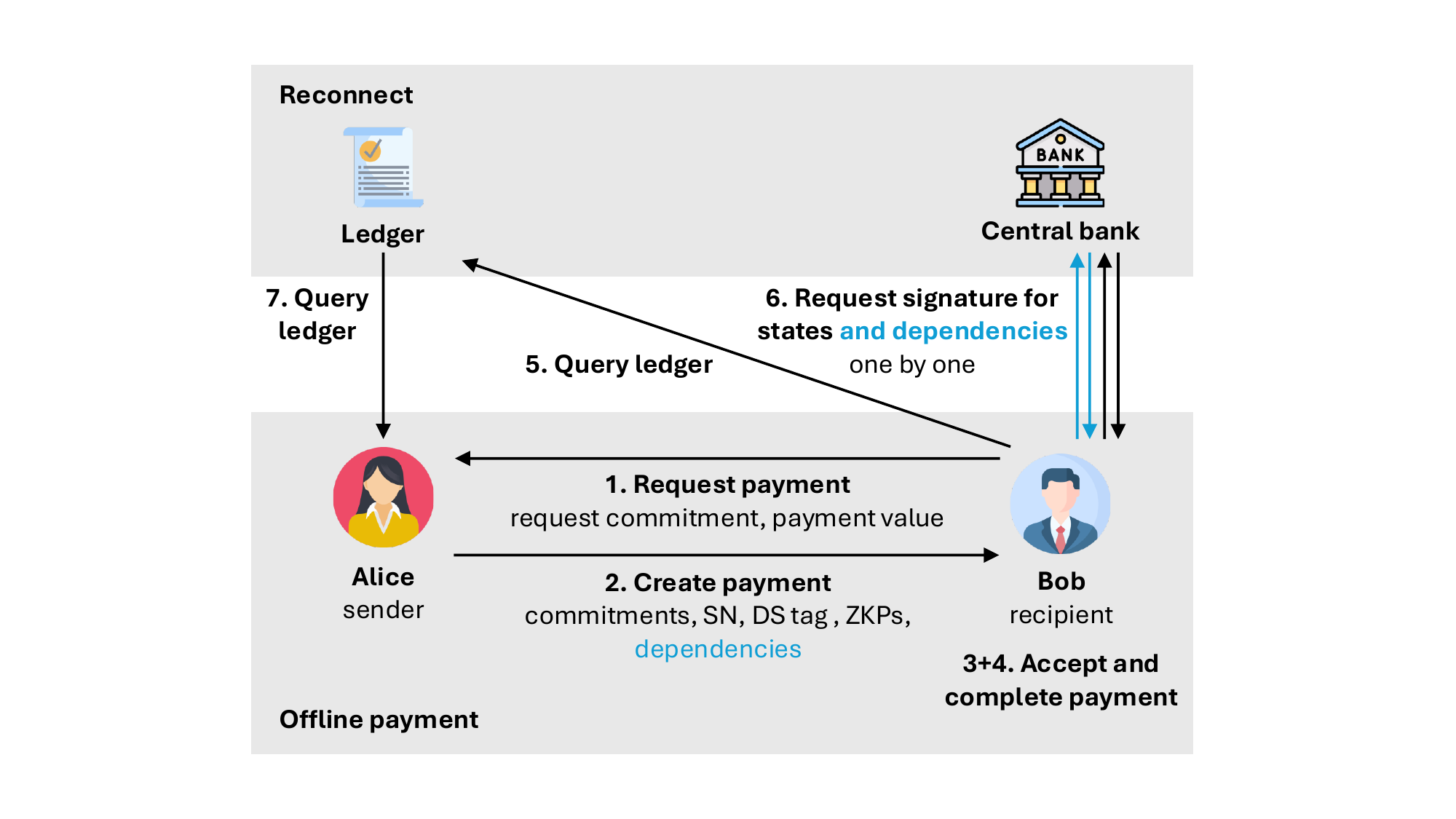}
    \caption{\name{} offline payment protocol. Elements marked in blue are only needed for multiple payments.}
    \vspace{-10pt}
    \label{fig:PaymentCon}
\end{figure}

We start by presenting a protocol that allows users to send or receive \emph{only one} payment offline, as illustrated in \cref{fig:PaymentCon} without the blue elements.  

\para{Offline payment} We assume that both Alice and Bob have recently reconnected and are now offline. Alice and Bob hold the openings to their state commitments, which include their \textit{secret key}, \textit{account balance}, \textit{holding limit}, and a \textit{counter}. The counter represents the number of payments that the user already sent, and is used to detect double spends. In addition, each state commitment includes the \textit{previous state commitment} of the user, and a \textit{counterparty commitment} provided by the counterparty of the corresponding payment. %
A payment from Alice to Bob proceeds as follows:

\begin{enumerate}
    \item \textbf{Request payment:} Bob sends a payment request to Alice. This request includes a \textit{request commitment} to his state commitment that hides his state commitment from Alice, as well as a \textit{payment value}. 
    \item \textbf{Create payment:} Alice reduces her account balance by the payment value, adds Bob's request commitment as the counterparty commitment of her new account state, and increases her payment counter by one. She computes her new state commitment and generates a \textit{serial number} and \emph{double-spending tag}. 
    The serial number uniquely identifies Alice's payment with the new counter value. 
    The double-spending tag is computed over Alice's new state commitment and counter value. The tag ensures that Alice can be identified if she double spends (makes two payments with the same serial number, but different state commitments). 
    To conclude her state transition, Alice generates a \textit{state transition ZKP} to prove that the values were updated correctly, as described above, and that her previous state was signed. 
    
    Next, Alice generates values that allow Bob to prove to the central bank that he received a payment from Alice, and thus that he is allowed to increase his balance. These are (i) a \textit{payment commitment} to the payment value, to Bob's request commitment, and to her new state commitment, and (ii) a \textit{payment ZKP} that shows that her new state resulted from her payment to Bob. Finally, Alice sends Bob her new state commitment, serial number, double spending tag, and state transition ZKP, alongside the openings of the payment commitment and its ZKP.  
    \item \textbf{Accept payment:} Bob verifies the received ZKPs, checks whether the payment commitment matches the expected values, and accepts the payment.
    \item \textbf{Complete payment:} Bob updates his state by increasing his balance and setting the counterparty commitment to Alice's new state commitment. Bob computes his new state commitment. For his ZKP, Bob needs to show (i) that all values were computed correctly and are consistent with Alice's payment, (ii) that his previous state is signed, and (iii) that Alice's new state is signed. However, since Alice's new state is not yet signed, Bob cannot create his ZKP yet. 
\end{enumerate}

\para{Reconnect} When Alice and Bob go back online, they reconnect to the central bank. Assume that Bob does this first. 

\begin{enumerate}
    \item[5)] \textbf{Query ledger:} Bob checks the ledger for existing signatures over Alice's new state commitment. Alice's state is not signed because Alice has not yet reconnected.
    \item[6)] \textbf{Request signatures:} Bob sends Alice's new state commitment, serial number, double spending tag, and ZKP to the central bank to request a signature for Alice's state. The central bank checks whether Alice's payment creation is a double spend by searching the ledger for an entry with the same serial number and a different state commitment. If Alice did not double spend and her ZKP verifies, the central bank signs her new state commitment, adds it to the ledger, and sends the signature to Bob. 

    Bob can now generate the missing ZKP for his own state and send a signature request consisting of his new state commitment, the payment commitment, and the state transition and payment ZKPs to the central bank. The central bank verifies all proofs, signs the state commitment, adds it to the ledger, and returns the new signature to Bob. For payment completions, the central bank does not need to check for double spends. 
\end{enumerate}

At this point, Bob is ready to go offline again and send or receive another payment. Finally, Alice reconnects.

\begin{enumerate}
    \item[7)] \textbf{Query ledger and download signature:} Alice checks the ledger and learns that her latest state commitment is already signed. She downloads the signature and is now ready to send or receive the next offline payment. 
\end{enumerate}

\begin{figure}[t]
    \centering
    \includegraphics[clip, trim=2.5cm 3.7cm 3.3cm 3.7cm, width=\linewidth]{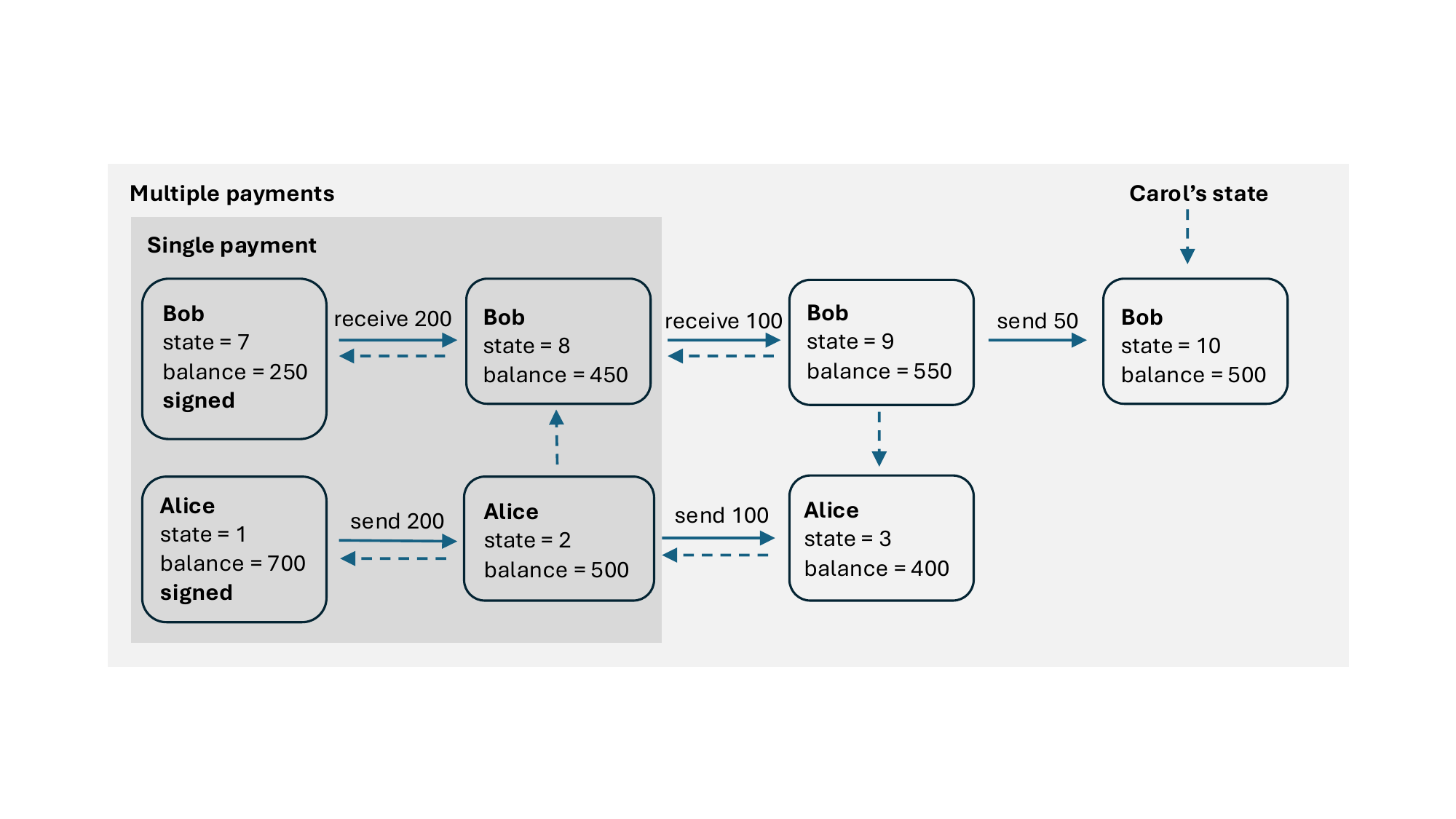}
    \caption{Evolution of user states across payments. Solid arrows indicate state transitions, dashed arrows indicate dependencies.}
    \vspace{-10pt}
    \label{fig:dependencies}
\end{figure}

\para{Dependencies of states} The ZKPs for Alice's and Bob's new states prove that their respective previous states are signed. We therefore call the previous state a \emph{dependency} of the new state. Alice's state has no further dependencies, her payment is valid even if Bob never receives it. 
In contrast, 
Bob can only receive his payment after Alice's payment creation state is signed. His payment completion state therefore also has a dependency on Alice's payment creation state, as shown in \cref{fig:dependencies}.

\subsection{Extension for Multiple Offline Payments}

We now extend this protocol to support multiple offline payments. 
In our previous protocol, Alice was able to create the ZKP for her payment state because her last state was signed. 
The same approach is not guaranteed to work, if users can engage in multiple payments offline, since Alice's state before her payment may not yet have a signature.

We address this problem by moving part of the state transition ZKP to a \textit{dependency ZKP}: The state transition ZKP now shows that the new state is computed correctly with respect to its dependencies (i.e., the previous state and, for payment completions, the sender's state). The \textit{dependency ZKP} proves that all dependencies of the state are signed. For an offline payment, Alice and Bob can generate the state transition ZKP, even if they may not yet have signatures for all dependencies. If Alice cannot yet generate a dependency ZKP, she needs to equip Bob with the necessary data to get any missing signatures to allow him to compute the dependency ZKPs for her state. To achieve this without compromising the confidentiality of Alice's state, we complement each state with a commitment to its dependencies. Alice can now share the openings of this \textit{dependency commitment} and the necessary data for her previous states when she pays Bob.

\cref{fig:dependencies} illustrates an example in which both Alice and Bob previously engaged in a payment while offline.
For Alice's next offline payment to Bob (state 3), Alice sends the payment commitment and ZKP and her state commitment, serial number, double spending tag, and state transition ZKP. Since Alice cannot generate the dependency ZKP yet, she shares the openings of her dependency commitment, which include the state commitment of her previous state 2. To allow Bob to get a signature over state 2 (which was also a payment creation), she also sends its state commitment, serial number, double spending tag, state transition ZKP, and dependency commitment and ZKP. This allows Bob to verify that Alice's new state resulted from a sequence of valid state transitions ($1 \rightarrow 2$, $2 \rightarrow 3$).

After receiving the payment from Alice, Bob makes an offline payment to Carol. Since Bob's current state 9 now depends on Alice's state 3 (which may be a double spend), Bob needs to include data for Alice's states 3 and 2 to pay Carol. Bob also includes data for his states 7, 8 and 9. 

When users reconnect, they anonymously upload their dependencies to the central bank, one by one. Users may add randomized delays between uploads to prevent time-based correlation (see \cref{sec:security}).
The central bank checks the correctness of each state transition (\property{S2}) and signs each state commitment. If a state is a double spend, the central bank observes the matching serial numbers and reconstructs the identity of the double spender from the double spend tags (\property{A2}).

\para{Offline payment} Our extended solution is illustrated in \cref{fig:PaymentCon}, where blue elements show additions to the previous protocol.

\begin{enumerate}
    \item \textbf{Request payment:} Bob computes a request commitment and sends it, together with the payment value, to Alice.
    \item \textbf{Create payment:} Alice updates her state, but compared to the single payment case, she additionally creates a dependency commitment and separates the dependency ZKP from the state transition ZKP. Alice sends her new state commitment, the openings of the payment commitment, and the state transition and payment ZKPs to Bob. She also includes either the required data about her state dependencies along with the openings of her dependency commitment, or the dependency commitment and the dependency ZKP.
	\item \textbf{Accept payment:} Bob verifies that Alice's state was computed correctly and that it has a valid dependency ZKP or that he has all data to generate one, when he reconnects. If all checks pass, he accepts the payment. 
	\item \textbf{Complete payment:} Bob updates his state and additionally creates a dependency commitment and a state transition ZKP. 
\end{enumerate}

\para{Reconnect} Assume that Bob reconnects first. 

\begin{enumerate}
    \item[5)] \textbf{Query ledger:} Bob first checks whether any of his own or his dependency states already has a signature. We assume that neither Alice nor Carol have reconnected yet, so this is not the case.
    \item[6)] \textbf{Request signatures:} In the example in \cref{fig:dependencies}, Bob first sends a signature request for his dependency state 2 from Alice. The central bank verifies that Alice's state transition (1 $\to$ 2) was computed correctly, adds her signed state (2) to the ledger, and returns the signature to Bob. Bob generates the missing ZKP for his dependency on Alice's state 3, and sends another signature request to the central bank. 
    Bob repeats the same steps to iteratively send signature requests for his states 8, 9, and finally 10. 
    If none of the payments was a double spend, Bob now has a signature over his most recent state 10.
\end{enumerate}

Alice reconnects later. 

\begin{enumerate}
	\item[7)] \textbf{Query ledger and download signature:} Alice checks the ledger and notices that her current state is already signed by the central bank, and downloads the signature. 
\end{enumerate}

\para{State recovery}
If Alice double spent in the example in \cref{fig:dependencies}, the central bank detects this when Carol uploads her dependencies when reconnecting. The central bank does not sign Alice's state, and therefore, Carol cannot create the necessary dependency ZKPs for Bob's and her states. However, since offline settlement is guaranteed, Carol can collect her money by executing \textit{state recovery}. To recover, Carol must identify herself and upload all her dependencies at once to the central bank to prove that she was a (potentially indirect) recipient of the fraudulent payment (\property{A3}). The central bank then signs Carol's state, and she can continue getting signatures for other unsigned states. 
Note that Carol would not have to recover her state if Bob had already executed state recovery: In this case, Bob's state would be signed, which allows Carol to generate the necessary dependency ZKP for her own state.

\para{Synchronization} Users must periodically extend the validity of their wallet by running an interactive synchronization protocol with the central bank. The central bank defines the interval after which the user must synchronize in discrete epochs. To enforce this, we include the last synchronization time (in epochs) in the account state.
During synchronization, the users update the epoch value through a state transition and prove that their previous state is signed. Thereby, the synchronization mechanism ensures that double spend recipients cannot delay state recovery indefinitely.

\section{Solution Details}

In this section, we describe our solution in detail. State recovery and synchronization details are deferred to \cref{sec:add_details}, complementary pseudocode is provided in \cref{sec:sel_pseudocode}

\begin{figure*} [htb]
\noindent 
\begin{minipage}[t]{\linewidth} 
\begin{mdframed}
\footnotesize
\begin{multicols}{3} 
    \textbf{Payment creation dependency $\mathsf{zkp}^{\mathsf{dep}}_{\mathsf{create}}$}
    \begin{itemize}
        \item Public values $pk_{\mathsf{CB}}$, $\mathsf{dcm}^{\mathsf{sen}}_{l+1}$
        \item Secret values $\mathsf{scm}^{\mathsf{sen}}_{l}$, $\sigma^{\mathsf{sen}}_{l}$, $\mathsf{blind}^{\mathsf{dep}}_{l+1}$
        \begin{itemize}
            \item $\mathsf{dcm}^{\mathsf{sen}}_{l+1} = \mathsf{comm}_{\mathsf{blind}_{l+1}^{\mathsf{dep}}}(\mathsf{scm}^{\mathsf{sen}}_{l})$
            \item $\mathsf{Ver}(pk_{\mathsf{CB}}, \mathsf{scm}^{\mathsf{sen}}_{l}, \sigma^{\mathsf{sen}}_{l}) = \mathsf{True}$
        \end{itemize}
    \end{itemize}
    \textbf{Payment creation state transition $\mathsf{zkp}^{\mathsf{state}}_{\mathsf{create}}$}
    \begin{itemize}
        \item Public values $\mathsf{scm}^{\mathsf{sen}}_{l+1}$, $\mathsf{dcm}^{\mathsf{sen}}_{l+1}$, $\mathsf{sn}_{l+1}$, $\mathsf{ds}_{l+1}$
        \item Secret values $sk$, $H$, $\mathsf{ctr}_{l}$, $\mathsf{bal}_{l}$, $e_{l}$, $v^{\mathsf{pm}}$, $\mathsf{scm}^{\mathsf{sen}}_{l-1}$, $\mathsf{ccm}_{l}$, $\mathsf{ccm}_{l+1}$, $\mathsf{blind}^{\mathsf{state}}_{l}$, $\mathsf{blind}^{\mathsf{state}}_{l+1}$, $\mathsf{blind}_{l+1}^{\mathsf{dep}}$
        \begin{itemize}
            \item $\mathsf{id}^{\mathsf{sen}}$ := $\mathsf{PRF}^{\mathsf{id}}_{sk}(0)$
            \item $\mathsf{scm}^{\mathsf{sen}}_{l}$ := $\mathsf{comm}_{\mathsf{blind}^{\mathsf{state}}_{l}}$($sk$, $H$, $\mathsf{ctr}_{l}$, $\mathsf{bal}_{l}$, $e_{l}$, $\mathsf{scm}^{\mathsf{sen}}_{l-1}$, $\mathsf{ccm}_{l}$)
            \item $\mathsf{scm}^{\mathsf{sen}}_{l+1}$ = $\mathsf{comm}_{\mathsf{blind}^{\mathsf{state}}_{l+1}}$($sk$, $H$, $\mathsf{ctr}_{l}+1$, $\mathsf{bal}_{l}-v^{\mathsf{pm}}$, $e_{l}$, $\mathsf{scm}^{\mathsf{sen}}_{l}$, $\mathsf{ccm}_{l+1}$)
            \item $\mathsf{sn}_{l+1}$ = $\mathsf{PRF}^{\mathsf{sn}}_{sk}(\mathsf{ctr}_{l} + 1 )$
            \item $\mathsf{ds}_{l+1}$ = $\mathsf{id}^{\mathsf{sen}} + \mathsf{scm}^{\mathsf{sen}}_{l+1} \cdot \mathsf{PRF}_{sk}^{\mathsf{ds}}(\mathsf{ctr}_{l} + 1)$ 
            \item $\mathsf{dcm}^{\mathsf{sen}}_{l+1}$ = $\mathsf{comm}_{\mathsf{blind}_{l+1}^{\mathsf{dep}}}(\mathsf{scm}^{\mathsf{sen}}_{l})$
            \item $\mathsf{bal}_{l} \geq v^{\mathsf{pm}}$
        \end{itemize}
    \end{itemize}\vfill\null \columnbreak
        \textbf{Payment completion dependency $\mathsf{zkp}^{\mathsf{dep}}_{\mathsf{comp}}$}
    \begin{itemize}
        \item Public values $pk_{\mathsf{CB}}$, $\mathsf{dcm}^{\mathsf{rec}}_{k+1}$
        \item Secret values $\mathsf{scm}^{\mathsf{rec}}_{k}$, $\mathsf{ccm}_{k+1}$, $\sigma^{\mathsf{rec}}_{k}$, $\sigma^{cp}_{k+1}$, $\mathsf{blind}^{\mathsf{dep}}_{k+1}$
        \begin{itemize}
            \item $\mathsf{dcm}^{\mathsf{rec}}_{k+1}$ = $\mathsf{comm}_{\mathsf{blind}_{k+1}^{\mathsf{dep}}}$($\mathsf{scm}^{\mathsf{rec}}_{k}$, $\mathsf{ccm}_{k+1}$)
            \item $\mathsf{Ver}(pk_{\mathsf{CB}}, \mathsf{scm}^{\mathsf{rec}}_{k}, \sigma^{\mathsf{rec}}_{k}) = \mathsf{True}$
            \item $\mathsf{Ver}(pk_{\mathsf{CB}}, \mathsf{ccm}_{k+1}, \sigma^{\mathsf{cp}}_{k+1}) = \mathsf{True}$
        \end{itemize}
    \end{itemize}
           \textbf{Payment completion state transition $\mathsf{zkp}^{\mathsf{state}}_{\mathsf{comp}}$}
    \begin{itemize}
        \item Public values $\Delta_{\mathsf{sync}}$, $\mathsf{scm}^{\mathsf{rec}}_{k+1}$, $\mathsf{dcm}^{\mathsf{rec}}_{k+1}$, $\mathsf{pcm}^{\mathsf{sen}}_{l+1}$
        \item Secret values $sk$, $H$, $\mathsf{ctr}_{k}$, $\mathsf{bal}_{k}$, $e_{k}$, $e^{\mathsf{sen}}_{l}$, $v^{\mathsf{pm}}$, $\mathsf{scm}^{\mathsf{rec}}_{k-1}$, $\mathsf{ccm}_{k}$, $\mathsf{ccm}_{k+1}$, $\mathsf{blind}^{\mathsf{req}}_{k}$, $\mathsf{blind}^{\mathsf{state}}_{k}$, $\mathsf{blind}^{\mathsf{state}}_{k+1}$, $\mathsf{blind}_{k+1}^{\mathsf{dep}}$, $\mathsf{blind}^{\mathsf{pm}}_{l+1}$
        \begin{itemize}
            \item $\mathsf{scm}^{\mathsf{rec}}_{k}$ := $\mathsf{comm}_{\mathsf{blind}^{\mathsf{state}}_{k}}$($sk$, $H$, $\mathsf{ctr}_{k}$, $\mathsf{bal}_{k}$, $e_{k}$, $\mathsf{scm}^{\mathsf{rec}}_{k-1}$, $\mathsf{ccm}_{k}$)
            \item $\mathsf{rcm}^{\mathsf{rec}}_{k}$ := $\mathsf{comm}_{\mathsf{blind}^\mathsf{req}_{k}}$($\mathsf{scm}^{\mathsf{rec}}_{k}$)
            \item $\mathsf{scm}^{\mathsf{rec}}_{k+1}$ = $\mathsf{comm}_{\mathsf{blind}^{\mathsf{state}}_{k+1}}$($sk$, $H$, $\mathsf{ctr}_{k}$, $\mathsf{bal}_{k}+v^{\mathsf{pm}}$, $e_{k}$, $\mathsf{scm}^{\mathsf{rec}}_{k}$, $\mathsf{ccm}_{k+1})$
            \item $\mathsf{pcm}^{\mathsf{sen}}_{l+1}$ = $\mathsf{comm}_{\mathsf{blind}_{l+1}^{\mathsf{pm}}}$($v^{\mathsf{pm}}$, $\mathsf{rcm}^{\mathsf{rec}}_{k}$, $\mathsf{ccm}_{k+1}$, $e^{\mathsf{sen}}_{l}$) 
            \item $\mathsf{dcm}^{\mathsf{rec}}_{k+1}$ = $\mathsf{comm}_{\mathsf{blind}_{k+1}^{\mathsf{dep}}}$($\mathsf{scm}^{\mathsf{rec}}_{k}$, $\mathsf{ccm}_{k+1}$)
            \item $\mathsf{bal}_{k} + v^{\mathsf{pm}} \leq H$
            \item $|e^{\mathsf{sen}}_{l} - e_{k}| \leq \Delta_{\mathsf{sync}}$
        \end{itemize}
    \end{itemize}\vfill\null \columnbreak
            \textbf{Payment $\mathsf{zkp}^{\mathsf{pm}}$}
    \begin{itemize}
        \item Public value $\mathsf{pcm}^{\mathsf{sen}}_{l+1}$
        \item Secret values $sk$, $H$, $\mathsf{ctr}_{l}$, $\mathsf{bal}_{l}$, $e_{l}$, $v^{\mathsf{pm}}$, $\mathsf{scm}^{\mathsf{sen}}_{l-1}$, $\mathsf{scm}^{\mathsf{sen}}_{l+1}$, $\mathsf{ccm}_{l}$, $\mathsf{ccm}_{l+1}$, $\mathsf{blind}^{\mathsf{state}}_{l}$, $\mathsf{blind}^{\mathsf{state}}_{l+1}$, $\mathsf{blind}^{\mathsf{pm}}_{l+1}$
        \begin{itemize}
            \item $\mathsf{scm}^{\mathsf{sen}}_{l}$ := $\mathsf{comm}_{\mathsf{blind}^{\mathsf{state}}_{l}}$($sk$, $H$, $\mathsf{ctr}_{l}$, $\mathsf{bal}_{l}$, $e_{l}$, $\mathsf{scm}^{\mathsf{sen}}_{l-1}$, $\mathsf{ccm}_{l})$
            \item $\mathsf{scm}^{\mathsf{sen}}_{l+1}$ = $\mathsf{comm}_{\mathsf{blind}^{\mathsf{state}}_{l+1}}$($sk$, $H$, $\mathsf{ctr}_{l}+1$, $\mathsf{bal}_{l}-v^{\mathsf{pm}}$, $e_{l}$, $\mathsf{scm}^{\mathsf{sen}}_{l}$, $\mathsf{ccm}_{l+1}$)
            \item $\mathsf{pcm}^{\mathsf{sen}}_{l+1}$ = $\mathsf{comm}_{\mathsf{blind}_{l+1}^{\mathsf{pm}}}$($v^{\mathsf{pm}}$, $\mathsf{ccm}_{l+1}$, $\mathsf{scm}^{\mathsf{sen}}_{l+1}$, $e_{l})$ 
        \end{itemize}
    \end{itemize}
        \textbf{Enrollment $\mathsf{zkp}^{\mathsf{enroll}}$}
      \begin{itemize}
        \item Public values $\mathsf{id}^{\mathsf{user}}$, $\mathsf{scm}^{\mathsf{user}}_{0}$,  $e$, $H$, $c$
        \item Secret values $sk$, $\mathsf{blind}_{0}^{\mathsf{state}}$ 
        \begin{itemize}
            \item $\mathsf{scm}^{\mathsf{user}}_{0}$ = $\mathsf{comm}_{\mathsf{blind}_{0}^{\mathsf{state}}}$($sk$, $H$, $0$, $e$, $0$, $c$)
            \item $\mathsf{id}^{\mathsf{user}}$ := $\mathsf{PRF}^{\mathsf{id}}_{sk}(0)$ 
        \end{itemize}
    \end{itemize}
  \end{multicols}
  \end{mdframed}
\end{minipage}
    \caption{Zero knowledge proofs for payment creation, payment completion, and user enrollment.}
    \vspace{-10pt}
    \label{fig:zkp_main}
\end{figure*}

\subsection{System Setup and State Representation}

To set up the CBDC system, the central bank generates a signature key pair $(sk_{CB}, pk_{CB})$, defines the security parameters, and performs the setup of the ZKP system.

The user's state consists of the following values. The user-chosen secret key $sk$ and the holding limit $H$ are fixed during enrollment. $\mathsf{ctr}_{i}$ counts how many payments a user has sent, and $\mathsf{bal}_{i}$ captures the current account balance. $e_{i}$ reflects the last epoch (a fixed time span determined by the central bank) in which the user synchronized. 
$\mathsf{scm}^{\mathsf{user}}_{i-1}$ represents the previous state commitment of the user and $\mathsf{ccm}_{i}$ represents the commitment to the state of the counterparty. The user commits to their current state by picking $\mathsf{blind}_{i}^{\mathsf{state}}$ randomly and computing the state commitment: $\mathsf{scm}^{\mathsf{user}}_{i} = \mathsf{comm}_{\mathsf{blind}_{i}^{\mathsf{state}}}(sk, H, \mathsf{ctr}_{i}, \mathsf{bal}_{i}, e_{i}, \mathsf{scm}^{\mathsf{user}}_{i-1}, \mathsf{ccm}_{i})$.

The user also maintains data structures called \textit{histories}, which contain past payment data indexed by the corresponding state commitment. The \textit{internal history} $\mathsf{hist}_{\mathsf{int}}$ holds the openings for the state commitment. The \textit{external history} $\mathsf{hist}_{\mathsf{ext}}$ stores data which the user needs to request signatures either (i) for their own states or (ii) for states from other users that the user received as dependencies. While internal history data remains private, external history data may be forwarded. The \textit{state recovery history} $\mathsf{hist}_{\mathsf{recovery}}$ holds the openings of the payment commitment for payment completions. 

\subsection{User Enrollment}

The user requests the current epoch $e$ and a challenge $c$ from the central bank. The user then selects a random secret key $sk$ and blinding value $\mathsf{blind}_{0}^{\mathsf{state}}$ to generate the initial state commitment $\mathsf{scm}^{\mathsf{user}}_{0}=\mathsf{comm}_{\mathsf{blind}_{0}^{\mathsf{state}}}(sk, H, 0, e, 0, c)$. The previous state commitment and account balance values are set to 0. $\mathsf{ccm}_{0}$ is set to the central bank challenge $c$ to ensure that all enroll requests are fresh. Additionally, the user derives the value $\mathsf{id}^{\mathsf{user}}$ pseudorandomly from the secret key, which acts as their \emph{identifier} towards the central bank. Lastly, the user generates the enroll ZKP $\mathsf{zkp^{enroll}}$
(see \cref{fig:zkp_main}). The user sends $\mathsf{scm}^{\mathsf{user}}_{0}$, $\mathsf{id}^{\mathsf{user}}$, $H$, $e$, and $c$ to the central bank, and adds the openings of the state commitment to $\mathsf{hist}_{\mathsf{int}}[\mathsf{scm}^{\mathsf{user}}_{0}]$.

The central bank verifies $\mathsf{zkp^{enroll}}$, and checks that $\mathsf{id}^{\mathsf{user}}$ is not already registered in their \textit{user registry} $R$. %
The central bank then adds $\mathsf{id}^{\mathsf{user}}$ to $R$ and %
approves the enrollment request by generating a signature $\sigma^{\mathsf{user}}_{0}$ over the state commitment $\mathsf{scm}^{\mathsf{user}}_{0}$ and returns the signature to the user. The central bank also stores $(\mathsf{scm}^{\mathsf{user}}_{0}, \sigma^{\mathsf{user}}_{0})$ in their \textit{ledger} $L$.
The user completes the enrollment by adding $\sigma^{\mathsf{user}}_{0}$ to $\mathsf{hist}_{\mathsf{ext}}[\mathsf{scm}^{\mathsf{user}}_{0}]$.

\subsection{Offline Payment}
\begin{figure*}[htb]
    \centering
    \includegraphics[clip, trim=0.7cm 21.2cm 8.7cm 1cm, width=0.9\textwidth]{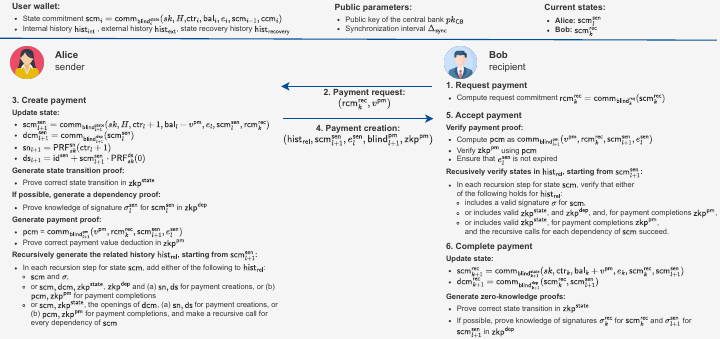}
    \caption{Offline payment protocol between the payment sender, Alice, and the payment recipient, Bob.}
    \vspace{-10pt}
    \label{fig:payment-protocol}
\end{figure*}

An offline payment proceeds as illustrated in \cref{fig:payment-protocol}.

\para{(1-2) Request payment}
The recipient sends a \emph{payment request} to the sender consisting of the payment value $v^{\mathsf{pm}}$ and the request commitment $\mathsf{rcm}^{\mathsf{rec}}_{k} := \mathsf{blind}_{k}^{\mathsf{req}}(\mathsf{scm}^{\mathsf{rec}}_{k})$. 

\para{(3-4) Create and send payment}
The sender creates a new payment state 
by decreasing the balance $\mathsf{bal}_{l}$ by the payment value $v^{\mathsf{pm}}$, and increasing the payment counter $\mathsf{ctr}_{l}$ by one. The user updates the previous state commitment to $\mathsf{scm}^{\mathsf{sen}}_{l}$ and the counterparty state commitment to $\mathsf{rcm}^{\mathsf{rec}}_{k}$. After choosing a random $\mathsf{blind}^{\mathsf{state}}_{l+1}$, the sender computes the new state commitment $\mathsf{scm}^{\mathsf{sen}}_{l+1}$ and stores its openings in $\mathsf{hist}_{\mathsf{int}}[\mathsf{scm}^{\mathsf{sen}}_{l+1}]$.
Using a random value $\mathsf{blind}^{\mathsf{dep}}_{l+1}$, the sender computes the state transition proof $\mathsf{zkp}^{\mathsf{state}}$ to show that the new state and dependency commitments, the double spending tag $\mathsf{ds}_{l+1}$ and the serial $\mathsf{sn}_{l+1}$ (see \cref{fig:zkp_main}) are computed correctly. If $\mathsf{scm}^{\mathsf{sen}}_{l}$ has a signature $\sigma^{\mathsf{sen}}_{l}$, the sender generates the dependency proof $\mathsf{zkp}^{\mathsf{dep}}$. The user stores $\mathsf{blind}^{\mathsf{dep}}_{l+1}, \mathsf{scm}^{\mathsf{sen}}_{l}$, $\mathsf{ds}_{l+1}$, $\mathsf{sn}_{l+1}$, and the computed ZKP(s) in the external history $\mathsf{hist}_{\mathsf{ext}}[\mathsf{scm}^{\mathsf{sen}}_{l+1}]$. 

The sender recursively generates the related history $\mathsf{hist}_{\mathsf{rel}}$. The related history either provides the signature for the new payment creation state, or the required data such that the recipient can obtain this signature when reconnecting.
In particular, the related history of state $\mathsf{scm}^{\mathsf{user}}_{i}$ consists of $\mathsf{scm}^{\mathsf{user}}_{i}$ and its signature $\sigma_{i}^{\mathsf{user}}$. If the sender does not yet have $\sigma_{i}^{\mathsf{user}}$, they instead include the required values for the signature request of $\mathsf{scm}_{i}^{\mathsf{user}}$ in the related history, i.e., $\mathsf{scm}^{\mathsf{user}}_{i},\mathsf{dcm}_{i}, \mathsf{zkp}^{\mathsf{state}}, \mathsf{zkp}^{\mathsf{dep}}$, and additionally $\mathsf{sn}_{i}$ and $\mathsf{ds}_{i}$ for payment creations or $\mathsf{pcm}_{i}$ and $\mathsf{zkp}^{\mathsf{pm}}$ for payment completions. If the sender also lacks $\mathsf{zkp}^{\mathsf{dep}}$ for $\mathsf{scm}_{i}^{\mathsf{user}}$, they add the openings of $\mathsf{dcm}_{i}$ to the related history instead of $\mathsf{zkp}^{\mathsf{dep}}$ and $\mathsf{dcm}_{i}$. In this case, to ensure that the recipient can generate $\mathsf{zkp}^{\mathsf{dep}}$ when reconnecting, the sender also adds the related history of each dependency of $\mathsf{scm}_{i}^\mathsf{{user}}$. %

Next, the sender picks a random $\mathsf{blind}_{l+1}^{\mathsf{pm}}$ to compute $\mathsf{pcm}_{l+1}^{\mathsf{sen}}=\mathsf{comm}_{\mathsf{blind}_{l+1}^{\mathsf{pm}}}(v^{\mathsf{pm}}, \mathsf{rcm}^{\mathsf{rec}}_{k}, \mathsf{scm}^{\mathsf{sen}}_{l+1}, e^{\mathsf{sen}}_{l})$ and generate the payment proof $\mathsf{zkp}^{\mathsf{pm}}$ 
(see \cref{fig:zkp_main}). Finally, the sender sends the openings of $\mathsf{pcm}^{\mathsf{sen}}_{l+1}$ together with $\mathsf{zkp}^{\mathsf{pm}}$, and the related history $\mathsf{hist}_{\mathsf{rel}}$ to the recipient, which concludes the offline payment on the sender side.

\para{(5) Accept payment}
The recipient checks that the sender's wallet is not expired, and computes $\mathsf{pcm}_{l+1}^{\mathsf{sen}}=\mathsf{comm}_{\mathsf{blind}_{l+1}^{\mathsf{pm}}}(v^{\mathsf{pm}}, \mathsf{rcm}^{\mathsf{rec}}_{k}, \mathsf{scm}^{\mathsf{sen}}_{l+1}, e^{\mathsf{sen}}_{l})$ to verify $\mathsf{zkp}^{\mathsf{pm}}$. 
Next, the recipient verifies that $\mathsf{hist}_{\mathsf{rel}}$ contains all the data to later request a signature over the sender's new state from the central bank. This check proceeds recursively and ensures that the sender correctly constructed the related history: Starting with $\mathsf{scm}^{\mathsf{sen}}_{l+1}$, the recipient verifies the respective ZKPs or the signature $\sigma$ for every $\mathsf{scm}$. If $\mathsf{hist}_{\mathsf{rel}}$ neither contains a dependency ZKP nor a signature $\sigma$, the recipient continues by checking all dependencies of $\mathsf{scm}$.
If all checks pass, %
the recipient accepts the payment.

\para{(6) Complete payment}
The recipient creates a new payment completion state as a transition from their current state. The balance is increased by $v^{\mathsf{pm}}$, the counterparty state commitment is set to $\mathsf{scm}^{\mathsf{sen}}_{l+1}$, and the previous state commitment is set to $\mathsf{scm}^{\mathsf{rec}}_{k}$. The user chooses a random $\mathsf{blind}_{k+1}^{\mathsf{state}}$ to compute the new state commitment $\mathsf{scm}^{\mathsf{rec}}_{k+1}$ and stores its openings in $\mathsf{hist}_{\mathsf{int}}[\mathsf{scm}^{\mathsf{rec}}_{k+1}]$.
Since the new state has a dependency on the sender's payment state, the recipient stores all received information from the related history $\mathsf{hist}_{\mathsf{rel}}$ in the external state history $\mathsf{hist}_{\mathsf{ext}}$. This ensures that the recipient can provide all required data if they subsequently pay other users.

The recipient generates the ZKPs. Notable differences to a payment creation are that the dependency ZKP also needs to show that the state of the sender is signed. In addition, the state transition proof $\mathsf{zkp}^{\mathsf{state}}$ also uses $\mathsf{pcm}_{l+1}^{\mathsf{sen}}$ as public input. Together with $\mathsf{zkp}^{\mathsf{pm}}$, this ensures that the recipient's payment value $v^{\mathsf{pm}}$ is consistent with the sender's payment value (see \cref{fig:zkp_main}). Thus, the recipient also stores $\mathsf{zkp}^{\mathsf{pm}}$ in $\mathsf{hist}_{\mathsf{ext}}[\mathsf{scm}^{\mathsf{rec}}_{k+1}]$.
To enable possible later state recovery, the recipient stores the openings of $\mathsf{pcm}^{\mathsf{sen}}_{l+1}$ in $\mathsf{hist}_{\mathsf{recovery}}[\mathsf{scm}^{\mathsf{rec}}_{k+1}]$.

\subsection{Reconnect}
\begin{figure*}[htb]
    \centering
    \includegraphics[clip, trim=0.6cm 17.3cm 0.5cm 1cm, width=0.9\textwidth]{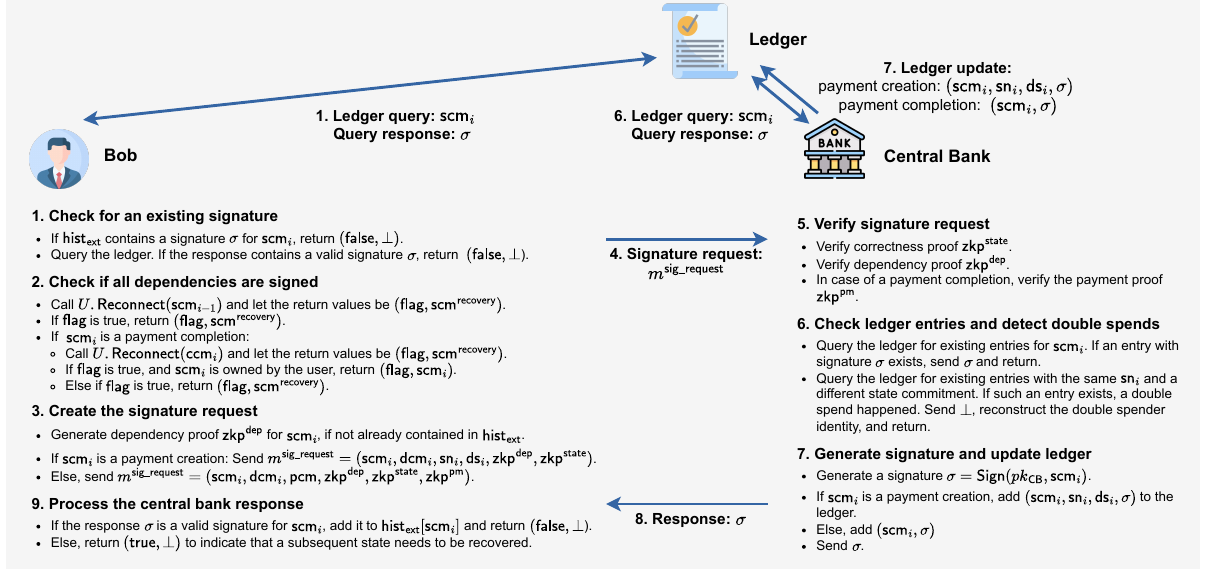}
    \caption{Users reconnect by recursively requesting signatures for unsigned states. The entry point is the most recent $\mathsf{scm}_{i}$.}
    \vspace{-10pt}
    \label{fig:reconnect}
\end{figure*}

Reconnect is shown in \cref{fig:reconnect}, and proceeds as follows. 

\para{1) Query ledger}
When a user reconnects, they need to get a signature over their most recent state. The user does this by \emph{recursively} requesting missing signatures over dependencies from their external history: The user identifies the newest unsigned state $\mathsf{scm}$ for which all dependencies are signed and checks the ledger $L$ for a signature $\sigma$ over $\mathsf{scm}$. 

\para{2)-4) Create and send signature request}
If $\sigma$ does not exist, the user creates $\mathsf{zkp}^{\mathsf{dep}}$ if needed. After that the user creates the signature request as $\mathsf{zkp}^{\mathsf{state}}$ and $\mathsf{zkp}^{\mathsf{dep}}$ with the respective public values for a payment state and $\mathsf{zkp}^{\mathsf{state}}$, $\mathsf{zkp}^{\mathsf{dep}}$, and $\mathsf{zkp}^{\mathsf{pm}}$ with the respective public values for a payment completion state and sends them to the central bank.

\para{5)-8) Processing and response to the signature request}
The central bank %
first verifies $\mathsf{zkp}^{\mathsf{state}}$, $\mathsf{zkp}^{\mathsf{dep}}$, and, for a payment completion, $\mathsf{zkp}^{\mathsf{pm}}$. 
Then, the central bank checks if the ledger $L$ contains a signature $\sigma$ for $\mathsf{scm}$ and responds with $\sigma$ if it does. For a payment creation, the central bank checks whether an entry with the same $\mathsf{sn}$ but a different $\mathsf{scm}$ exists, which would indicate that this payment creation is a double spend. In this case, the central bank reconstructs the identity of the double spender using the double spending tags and responds with $\bot$. %

Finally, if all checks pass, the central bank generates a signature $\sigma$ over $\mathsf{scm}$, adds an entry to the ledger $L$, and responds by sending $\sigma$ to the user.

\para{9) Process response}
If the user receives a signature $\sigma$, it updates $\mathsf{hist}_{\mathsf{ext}}[\mathsf{scm}]$ with $\sigma$. If the central bank rejected the payment, the user concludes that $\mathsf{scm}$ was a double spend and that they need to run state recovery.

\section{Security Analysis}
\label{sec:security}

In this section, we analyze the user privacy, payment security, and accountability guarantees of \name{}.

\subsection{Privacy Guarantees} 

\para{Payment and balance confidentiality (\property{P1})} 
Signature requests for payment creation states contain a state and dependency commitment $\mathsf{scm}$ and $\mathsf{dcm}$ which are hiding, a pseudo-random serial number $\mathsf{sn}$ and double spend tag $\mathsf{ds}$ which appear random if the sender did not double spend, and the zero-knowledge proofs $\mathsf{zkp}^{\mathsf{state}}$ and $\mathsf{zkp}^{\mathsf{dep}}$. Signature requests for payment completion states contain a state, dependency, and payment commitment $\mathsf{scm}$, $\mathsf{dcm}$, $\mathsf{pcm}$ and zero-knowledge proofs $\mathsf{zkp}^{\mathsf{state}}$, $\mathsf{zkp}^{\mathsf{dep}}$, $\mathsf{zkp}^{\mathsf{pm}}$. Since all uploaded values are indistinguishable from random values, identities, payment values, and account balances are confidential to the central bank.

A malicious central bank could falsely flag payments as double spends and trigger state recovery for a victim user. During state recovery, the user must disclose their identity and payment details to the central bank to continue using the system. Such attacks could be prevented by requiring the central bank to provide proof of double spending.\footnote{To present such a proof, the central bank would need to additionally store $\mathsf{zkp}^{\mathsf{state}}$ and $\mathsf{dcm}$ for each payment creation, which we consider acceptable storage overhead. In addition, PRFs need to additionally provide collision-resistance, and the system should be set up such that the central bank does not have access to the simulation trapdoor for the ZKP system, which we also consider reasonable assumptions for a CBDC system.} Our design and prototype could be extended to incorporate this solution.  

\para{Payment unlinkability (\property{P2})} Because all values that users send to the central bank appear random, the central bank cannot link (i) states of the same user or payment, or (ii) payments that involve the same sender or recipient, based on the content of the signature requests alone. Thus, all offline payments are cryptographically unlikable. However, in certain cases that we discuss next, the central bank may be able to leverage additional information, such as the time and number of signature requests or ledger queries, to infer payment links.

Recall that users may have multiple unsigned states when reconnecting, and therefore, they need to get signatures for multiple states, either through ledger queries or signature requests. Suppose that the central bank can correlate all signature requests from the same user (despite anonymous communication). In this case, it learns that all the correlated states are either dependencies on states from other users due to payment completions, or states from the reconnecting (anonymous) user. %
Combined with the fact that each signature request explicitly discloses the type of the state (payment creation or completion), this may help the central bank to de-anonymize such correlated states. %
The same leakage applies if the central bank can correlate anonymous ledger queries by the same user, e.g., due to their timing. 

We consider such correlation and de-anonymization attacks difficult to execute in practice: In a large CBDC deployment, many users are likely to request signatures simultaneously. Simple countermeasures, such as short randomized waits between signature requests and ledger accesses, would further complicate such attacks. In addition, PSPs could maintain replicas of the ledger and users could query those replicas. This would partially leak ledger access patterns to the PSPs but not to the central bank. Thus, inference attacks would require collusion between PSPs and the central bank.

In contrast, payments executed after a full reconnect are unlinkable to previous signature requests and ledger query sequences. In this case, the user knows the signature of their most recent state and can either create a dependency ZKP for the next state or forward its signature when creating payments. This prevents the above linking attacks because neither the user nor any direct or indirect recipient must request a signature or query the ledger for that state. 

We conclude that \name{} provides offline session unlinkability (\property{P2b}) for \emph{all} payments, whereas full payment unlinkability (\property{P2a}) is not guaranteed in all cases. 

\para{Payment privacy towards other users (\property{P3})} First, we consider the privacy of a payment recipient towards the sender. The payment creation message contains the sender's new state commitment, the related history, and openings for the payment commitment $\mathsf{pcm}$ and the payment proof $\mathsf{zkp}^{\mathsf{pm}}$
Since the sender knows $\mathsf{pcm}$ and $\mathsf{zkp}^{\mathsf{pm}}$, the sender can recognize the recipient's payment completion state if they receive a payment that includes the recipient's state in the related history.

Next, we consider the privacy of payment senders towards direct and indirect payment recipients. When the sender creates a payment, they disclose data related to the payment creation state, and possibly its dependencies, in the related history. The recipient relays this information when they subsequently pay other users. This allows users to recognize and link states that they receive in different payments. If users synchronize between payments, such links can be prevented. 

Finally, we consider a malicious user who colludes with the central bank. Such a user could pay a victim recipient and then help the central bank to identify the recipient's payment completion state based on the values $\mathsf{pcm}$ and $ \mathsf{zkp}^{\mathsf{pm}}$ in the signature request. Since these values are not stored on the ledger, the malicious user could not identify the recipient state independently, without the central bank. 

We conclude that in \name{}, the payments of a user from an offline session are not unlinkable to the payment counterparties in the same offline session. Towards other users and across offline sessions, payments are both confidential and unlinkable, if a user synchronizes when reconnecting. In \cref{subsec:unlinkability-wo-sync}, we present a modification to our protocol that ensures unlinkability of offline sessions without an additional synchronization step, which preserves offline synchronization despite collusion of the central bank and other users.

\para{Residual leakage} As the final part of our privacy analysis, we consider residual leakages that allow the central bank to infer something about offline payments, but not link payments to the same user or break payment confidentiality.

If the central bank operates the ledger, it learns the number of queries per state commitment. This number is a lower bound on the number of users that received this state as a dependency. 
Counting ledger queries is practical for the central bank even if there are many users and payments in the system (as opposed to correlation signature requests by the same user). However, it is hard to imagine how this information could be used to violate users' privacy. Such residual leakage is similar to transferable E-cash schemes, in which the size of the coin grows with each payment~\cite{chaum1992transferred}, which allows the bank to infer the number of offline payments for each coin at deposit time.

A malicious central bank could also infer a lower bound on the number of users who depend on a particular state by manipulating the ledger. One such strategy is to remove previously added signatures, and thus force every user to send a separate signature request to the central bank. Ledger manipulation attacks would be publicly visible to users and third parties, and we consider them unlikely in CBDC deployments.

\subsection{Security Guarantees}

\para{Payment security (\property{S1})} To create a valid payment, a user must prove knowledge of the openings of their previous state commitment in the state transition ZKP. Since the user never shares the secret key or the blind of the state commitment, only the rightful owner knows all required openings and can generate a valid ZKP without breaking the soundness of the ZKP system or the binding property of the commitment scheme. Therefore, \name{} provides payment security.

\para{Payment integrity (\property{S2})} Here, we consider a malicious user who has not compromised their device. Such a user can manipulate the communication channel, for example, by delaying the delivery of messages to the secure element, but they cannot arbitrarily deviate from the payment protocol. To create a payment without reducing the account balance, the user must create a payment that does not result from a sequence of state transitions starting from a signed state (which would violate simulation-extractability of ZKPs or binding of commitments) or forge a central bank signature (which would violate unforgeability of signatures). Completing the same payment more than once would require two openings for the same request commitment (which would violate the binding property of commitments) or forging a ZKP (violating simulation-extractability). Thus, payment integrity is guaranteed for users who have not compromised their secure elements.

\subsection{Accountability Guarantees}

\para{Holding limit integrity (\property{A1})} In an offline payment, the recipient initially sends a request commitment that the sender includes in the payment creation state. The recipient can only generate a valid state transition ZKP if the request commitment opens to the previous state commitment. Therefore, any state transition invalidates all other outstanding payments to the user\footnote{Users who compromised their secure element can request multiple parallel payments and complete them by creating multiple state transitions from the same state (similar to a double spend). However, if the user pays from more than one of those state transition sequences, this creates a double spend.}. This prevents the recipient from holding more money than captured in the account balance. Therefore, for holding limit integrity, it is sufficient to prove that each payment completion is compliant with the pre-defined holding limit.

\para{Counterfeit creator identification (\property{A2})} To identify a counterfeit creator, the central bank needs to receive two matching serial numbers with different state commitments. Since two correctly computed double spending tags with the different state commitments reveal the sender identity, a user can only evade detection if (i) the central bank never receives this data or (ii) they did not correctly compute the state commitment, serial number, or double spending tag. 

Once any direct or indirect recipient of the double spending reconnects, they send a signature request that includes the required data unless the state commitment is already included in the ledger, in which case the central bank has already received the double spending tag. Since the honest (non-colluding) recipient verified the state transition before accepting the payment, the data must have been correctly computed unless the sender can forge a central bank signature, break the simulation-extractability of the ZKP system, or break the binding property of the commitment scheme. 

\para{Counterfeit recipient de-anonymization (\property{A3})}
If a user reconnects and requests signatures for unsigned states, the central bank detects any payment creations that lead to counterfeits (see \property{A2}), and refuses to sign them. If a user received a payment with a direct or indirect dependency on the double spend state, the user cannot get a signature for their own state through a regular signature request. To get a signature through state recovery, the user must disclose their identity, and the received payment value, which satisfies the required recipient detection. If the user does not recover their state, and no previous recipient has recovered their state, the user has to eventually either stop using their account, discard the received payment, or become a double spender themselves due to the synchronization requirement (see \property{A4}). 

\para{Incomplete synchronization detection (\property{A4})}
When user synchronizes, they create a state transition from their most recent state and prove that this state is signed (\cref{sec:add_details}). If a user instead creates a transition from an older state, the impact is similar to that of a malicious recipient who completes two concurrent payments: The user can only pay from one of the two state transition sequences without causing a double spend. 

\section{Performance Evaluation}
\label{sec:performance-eval}

\subsection{Evaluation Setup}

To evaluate our solution, we implemented a non-optimized prototype in Go. We used the gnark~\cite{gnarklib} library to implement the zero-knowledge proofs using the Groth16~\cite{groth2016size} proof system in the BN254 group. Our implementation uses EdDSA~\cite{bernstein2012high} for signatures and MiMC~\cite{albrecht2016mimc} as hash function. Public key encryption uses ElGamal encryption, while pseudo-random functions and commitments are implemented using MiMC hashes. The size of keys, commitments, serial numbers, and double spending tags is 254 bits.
Evaluations were run on a laptop with an AMD Ryzen 7 PRO 7730U processor with 32GB RAM and averaged over 100 runs. Network effects were not measured, and data exchange was assumed to be instant. %

\begin{table} [htb]
\begin{center}
\begin{tabular}{@{}p{0.4\columnwidth} S[table-format=1.4] S[table-format=1.4]  S[table-format=5.0] @{}} 
 \toprule
\tablehead{0.4\columnwidth}{Proof Circuit} & \tablehead{0.2\columnwidth}{Generation (s)}& \tablehead{0.2\columnwidth}{Verification (s)} & \tablehead{0.2\columnwidth}{Constraints (\#)}
\\\midrule
Enroll  & 0.0239 & 0.0012 & 2312\\
Payment  & 0.0552 & 0.0013 & 6931\\
Payment creation state transition  & 0.0741 & 0.0013 & 9801\\
Payment creation dependency & 0.0512 & 0.0013 & 7662\\
Payment completion state transition  & 0.0795 & 0.0013 & 12674\\
Payment completion dependency  & 0.0906 & 0.0014 & 14993\\
Synchronization  & 0.0857 & 0.0014 & 12283\\
State recovery  & 0.0828 & 0.0014 & 11954\\
\bottomrule
\end{tabular}
\end{center}
\captionof{table}{Evaluation of zero-knowledge proof circuits.}
\vspace{-10pt}
\label{tab:eval_circ}
\end{table}

\subsection{User Operations} 

Subsequently, we report the results for the different user operations. In addition, \cref{tab:eval_circ} shows the proof generation and verification times as well as the number of R1CS constraints for each proof circuit used in \name{}.

\begin{table*}[htb]

\begin{center}
\begin{tabular}{@{}S[table-format=3.0] S[table-format=1.4] S[table-format=1.4] S[table-format=1.4] S[table-format=1.4] S[table-format=1.4] S[table-format=1.3] S[table-format=2.3] @{}} 
 \toprule
\multicolumn{2}{@{}c}{\textbf{Sender}} & \multicolumn{2}{c}{\textbf{Receiver}} & \multicolumn{2}{c}{\textbf{End to end}} & \multicolumn{2}{c@{}}{\textbf{Message sizes}}
\\ \cmidrule(lr){1-2} \cmidrule(lr){3-4} \cmidrule(lr){5-6} \cmidrule(lr){7-8}
\tablehead{0.125\textwidth}{Unsigned  history \newline (\# elem.)} &
\tablehead{0.125\textwidth}{Create Payment \newline (s)}  & 
\tablehead{0.125\textwidth}{Accept Payment \newline (s)} & 
\tablehead{0.125\textwidth}{Complete Payment \newline (s)} & 
\tablehead{0.125\textwidth}{Payment  accepted \newline (s)} & 
\tablehead{0.125\textwidth}{Payment  completed \newline (s)} & 
\tablehead{0.125\textwidth}{Request Payment (kB)} & 
\tablehead{0.125\textwidth}{Create Payment \newline (kB)}
\\\midrule
1 & 0.2097 & 0.004 & 0.0917 & 0.2138 & 0.3054 & 0.077 & 1.033\\
51 & 0.1554 & 0.0944 & 0.0903 & 0.2499 & 0.3401 & 0.077 & 46.990\\
101 & 0.157 & 0.1843 & 0.0903 & 0.3413 & 0.4317 & 0.077 & 92.945\\
\bottomrule
\end{tabular}
\end{center}
\captionof{table}{Evaluation of offline payments based on the sender's unsigned history size after the payment creation.}
\vspace{-10pt}
\label{tab:eval_payment}
\end{table*}

\para{Offline payment} First, we evaluate offline payments by varying the number of unsigned elements in the sender's external history. In \cref{tab:eval_payment}, we report payment creation, acceptance and completion times, as well as message sizes. Once a payment is \emph{accepted}, the sender can participate in other payments. Once a payment is \emph{completed}, the receiver can participate in other payments. We omit the payment request computation times, which were always below 0.1ms.

We note that a sender can only create a dependency ZKP if they have exactly one unsigned history element after the payment creation. In this case, the payment creation takes around 210ms, longer unsigned history sizes take around 50ms less. The execution time for payment acceptance increases linearly by around 90ms per 50 unsigned history elements. At 101 elements, the end-to-end time is 341ms until payment acceptance and 432ms until payment completion.

A payment request message is 77B and can be transferred using QR codes. The message size of payment creations increases linearly by around 46kB per 50 unsigned history elements. As at most 3kB can be stored on a QR code, only payment creations with up to two unsigned history elements could be sent via QR codes, making them unsuitable for this protocol. NFC connections have a transmission rate of around 420kbit/s. 
Combining the computation and NFC transmission times, the end-to-end time until payment acceptance takes around 1.1s at 51 unsigned history elements and around 2.2s at 101. This complies with the Digital Euro regulation which requires settlement in ``a matter of seconds''~\cite{euro-regulation}. For comparison, a typical Mastercard payment takes 630ms$\pm$150ms~\cite{coppola2024pure}.

\para{Reconnect}
\cref{tab:eval_recon} shows the time needed for a user to reconnect based on the number of unsigned elements in their history and the total size of messages that the user needs to send to the central bank. %
In the evaluated scenario, only one of the unsigned states already has a dependency ZKP. In this case, the reconnect time increases linearly by around 4.2s for every 50 unsigned history elements. This is because the user needs to compute a dependency ZKP for all other unsigned states. In practice, users may have more dependency ZKPs in their history or states may already have signatures, which would shorten the reconnect time. The total size of messages that need to be sent to the central bank during reconnect increases linearly by around 40kB per 50 unsigned history elements. As reconnecting is typically executed using WiFi or cellular connection, we do not consider this a problem.

\begin{table} [htb]
\begin{center}
\begin{tabular}{@{}S[table-format=3.0] S[table-format=1.4] S[table-format=2.3] @{}} 
 \toprule
\tablehead{0.3\columnwidth}{Unsigned history \newline (\# elem.)} & 
\tablehead{0.3\columnwidth}{Reconnect without \newline state recovery (s)} &
\tablehead{0.4\columnwidth}{Total size of messages sent to the central bank (kB)}
\\\midrule
1 & 0.0029 &  0.786\\
51 & 4.2891 & 42.286\\
101 & 8.5675 & 83.786\\
\bottomrule
\end{tabular}
\end{center}
\captionof{table}{Evaluation of reconnect based on the user's unsigned history size (without wait times).}
\vspace{-10pt}
\label{tab:eval_recon}
\end{table} 

\para{History sizes} To analyze history sizes, we distinguish between merchants, who primarily receive payments, and consumers, who primarily send payments. We focus on consumers because merchants are likely to reconnect regularly. We assume that (i) all consumers make 1.3 payments per day (the total number of credit card payments in the U.S.~\cite{uscredit} averaged over the U.S. population~\cite{uspopulation}), (ii) consumers receive payments from a group of users who never received payments, and (iii) all consumers stay offline equally long and start with no unsigned states in their history. 
We consider the following scenarios: A) Short internet or central bank outage of around one day. B) Consumers who choose to stay offline for a week up to a month. %
C) Consumers who do not have consistent connectivity and need to stay offline for multiple months.
We analyze the size of the unsigned history at the end of the offline period and the \textit{final payment time}, i.e., the end-to-end time until payment completion for the last offline payment using NFC.

\noindent\emph{Scenario A:} Consumers who receive one payment per day would have an unsigned history size of 3.9, the final payment would take roughly 340ms. Our system can handle such outages (and thus increase the financial system's stability).

\noindent\emph{Scenario B:} Consumers who stay offline for a week and receive payments daily would have an unsigned history size of 54.8 and a final payment time of around 1.31s. Consumers who stay offline for a month and receive payments monthly would have an expected unsigned history size of 60.3 and a final payment time of around 1.43s. %
Payments could be accelerated by using Bluetooth instead of NFC.

\noindent\emph{Scenario C:} If consumers are offline for six months, their history size would be 237.1 even if they do not receive payments; the final payment would take almost 5s. Thus, our system is not ideal for such users and scenarios.

\subsection{Central Bank Operations}
We report the evaluation results of the central bank operations in \cref{tab:eval_cb}. Furthermore, \cref{tab:eval_circ} shows the proof generation and verification times and the number of R1CS constraints for all zero-knowledge proof circuits used in \name{}. 

Processing signature requests takes 3ms for payment creations and 4ms for payment completions.
The latter operation is 1ms slower, since the payment proof must also be verified.

For every offline payment, the central bank must process two states: the sender's creation state and the recipient's completion state. The processing time of both signature requests combined is around 7ms if no double spend occurred. Thus, our system can reach a throughput of 143 payments per second using sequential execution. \name{} is amenable to parallel processing. Signature requests for payment completions can be parallelized, since they don't involve double spending checks. Signature requests for payment creations can be parallelized if serial numbers are queried sequentially. Serial numbers from distinct ranges could be queried in parallel from separate logs. In 2021, an average of around 5,000 credit card payments per second were completed in the U.S.~\cite{uscredit, uspopulation}. Using \name{} with parallel processing, 35 cores are needed to handle such a load. On days with 5 times higher payment volume, the central bank would need 175 cores, which we consider feasible.

\begin{table} [htb]
\begin{center}
\begin{tabular}{@{}p{0.6\columnwidth} S[table-format=1.4] @{}} 
 \toprule
\tablehead{0.6\columnwidth}{Central bank operation} & \tablehead{0.3\columnwidth}{Processing~time (s)}
\\\midrule
Enrollment request  & 0.0016\\
Signature request (Payment creation, no DS) & 0.0029\\
Signature request (Payment creation, DS)  & 0.0026\\
Signature request (Payment completion)  & 0.0041\\
State recovery request (without history verification)  & 0.0017\\
Synchronization request & 0.0017\\
\bottomrule
\end{tabular}
\end{center}
\captionof{table}{Execution time of central bank operations.}
\vspace{-10pt}
\label{tab:eval_cb}
\end{table}

\section{Discussion}
\label{sec:discussion}

\para{Money destruction}
In \name{}, money can be accidentally destroyed due to crash-failures either on the communication channel or of user wallets during payments, even if the failure is only temporary. Suppose a sender creates a payment, but the message does not reach the recipient. The message delivery can be re-tried, however, if the recipient executes another payment before receiving the previous message, the recipient cannot complete the payment anymore. 
To limit losses, it is possible to extend \name{} to allow senders to credit the payment value back to their account if (i) the recipient proves that they did not and will never complete the payment and (ii) the sender has not yet created another payment. (i) prevents counterfeit creation, whereas (ii) ensures that holding limits cannot be evaded. 
The central bank may decide to relax either of these requirements and, e.g., refund payments on a case-by-case basis if a user identifies themselves when going back online.

\para{Role of PSPs}
This paper focuses on the end-users and the central bank. However, in a CBDC deployment, the PSPs also play a significant role. In \name{}, PSPs could help minimize the amount of information that users need to send to the central bank. For example, PSPs could replicate the central bank's ledger, which addresses some residual leakages identified in \cref{sec:security}. Similarly, PSPs could improve privacy towards the central bank by assisting in the state recovery protocol. %

\para{Economic implications}
CBDCs as a new form of legal tender make the payment system less dependent on third parties and may help serve currently unbanked populations. However, since CBDCs have the potential to disrupt banking, care needs to be taken to avoid bank-runs in which users exchange their bank deposits for CBDCs, destabilizing the financial system in the process. Another risk is that CDBCs crowd out deposits and lead to an increase in funding costs for banks and a decrease in investments~\cite{keister2023}. Cash-like CBDCs with holding limits are less problematic because they do not compete with deposits. Under certain conditions, CBDCs may crowd out cash, notably if they provide sufficient privacy protection~\cite{agur2022}. \name{} is a cash-like CBDC with holding limits, and thus, avoids the risk of crowding out deposits. The crowding out of cash can be accepted or resolved via policy. 

In order to be operational and widely used, CBDCs should support  transferable offline payments~\cite{keister2023}. Offline payments can enable attackers to create counterfeit money that cannot be detected immediately as the recipient cannot query the central bank to ensure the received money is legitimate. Secure elements are often proposed to prevent attackers from creating counterfeit, however, secure elements can be broken. Thus, it is important to ensure that counterfeit and counterfeit creators can be detected to prevent counterfeit from flooding the system without the central bank knowing. This begs the question if users can exchange received counterfeit for real money or not. In most currently proposed systems they cannot, meaning recipients bear the risk and may lose their money. This is similar to cash where accepting a counterfeit bill may lead to losing money. However, this approach may not be acceptable for a CBDC: In traditional cash, a bill can be inspected and identified as counterfeit, but in CBDC systems, honest users have no chance to identify counterfeit. This may make many hesitant to accept CBDC. However, allowing recipients to exchange their counterfeit for legitimate money, such as in \name{}, could open up many complex collusion patterns that need to be investigated and dealt with. Furthermore, the creation and distribution of large amounts of counterfeit could destabilize the system. Thus, robust policies and fallback methods need to be defined to handle these cases. One potential solution for counterfeit becoming a serious and undetected problem is to introduce requirements for users to regularly synchronize with the central bank. If a wallet is not synchronized in time, it expires and cannot be used until synchronized.

\section{Conclusion}
\label{sec:conclusion}

Anonymous payments and privacy-preserving digital currencies have been studied for decades, and various schemes exist in the research literature. However, recent regulation proposals for CBDCs present new requirements and a different risk model, which have not been considered previously. To the best of our knowledge, this paper is the first to design an offline payment solution that combines strong privacy protection with expressive regulation features, such as holding limits, in a way that is robust to secure element failure.

\bibliographystyle{plain}
\bibliography{references}

\appendix
\subsection{Additional Solution Details}
\label{sec:add_details}

\begin{figure*} [htb]
\noindent 
\begin{minipage}[t]{\linewidth} 
\begin{mdframed}
\small
\begin{multicols}{2}
        \textbf{State Recovery ZKP $\mathsf{zkp}^{\mathsf{recovery}}$}
   \begin{itemize}
        \item Given public values $pk_{\mathsf{CB}}$, $\mathsf{id}^{\mathsf{user}}$, $v^{\mathsf{pm}}$, $\mathsf{scm}^{\mathsf{user}}_{i}$, $\mathsf{pcm}$, 
        \item I know secret values $sk$, $H$, $\mathsf{ctr}_{i}$, $\mathsf{bal}_{i}$, $e^{\mathsf{user}}_{i}$, $e^{\mathsf{sen}}$, $v^{pm}$, $\mathsf{scm}^{\mathsf{user}}_{i-1}$, $\mathsf{ccm}_{i}$, $\mathsf{blind}^{\mathsf{state}}_{i}$, $\mathsf{blind}_{i-1}^{\mathsf{req}}$, $\mathsf{blind}^{\mathsf{pm}}$, $\sigma^{\mathsf{user}}_{i-1}$
        \item such that 
        \begin{itemize}
            \item $\mathsf{rcm}^{\mathsf{user}}_{i-1}$ := $\mathsf{comm}_{\mathsf{blind}_{i-1}^{\mathsf{req}}}$($\mathsf{scm}^{\mathsf{user}}_{i-1}$)
            \item $\mathsf{scm}^{\mathsf{user}}_{i}$ = $\mathsf{comm}_{\mathsf{blind}_{i}^{\mathsf{state}}}$($sk$, $H$, $\mathsf{ctr}_{i}$, $\mathsf{bal}_{i}$, $e^{\mathsf{user}}_{i}$, $\mathsf{scm}^{\mathsf{user}}_{i-1}$, $\mathsf{ccm}_{i}$)
            \item $\mathsf{id}^{\mathsf{user}}= \mathsf{PRF}^{\mathsf{id}}_{sk}(0)$
            \item $\mathsf{Ver}(pk_{\mathsf{CB}}, \mathsf{scm}_{i-1}^{\mathsf{user}}, \sigma^{\mathsf{user}}_{i-1}) = \mathsf{True}$ %
            \item $\mathsf{pcm}$ = $\mathsf{comm}_{\mathsf{blind}^{\mathsf{pm}}}$($v^{\mathsf{pm}}$, $\mathsf{rcm}^{\mathsf{user}}_{i-1}$, $\mathsf{ccm}_{i}$, $e^{\mathsf{sen}}$)
        \end{itemize}
    \end{itemize}\vfill\null \columnbreak
        \textbf{Synchronization ZKP $\mathsf{zkp}^{\mathsf{sync}}$}
    \begin{itemize}
        \item Given public values $pk_{\mathsf{CB}}$, $\mathsf{scm}^{\mathsf{user}}_{i+1}$, $e$, $c$
        \item I know secret values $sk$, $H$, $\mathsf{ctr}_{i}$, $\mathsf{bal}_{i}$, $e_{i}$, $\mathsf{scm}^{\mathsf{user}}_{i-1}$, $\mathsf{ccm}_{i}$, $\mathsf{blind}^{\mathsf{state}}_{i}$, $\mathsf{blind}^{\mathsf{state}}_{i+1}$, $\sigma^{\mathsf{user}}_{i}$
        \item such that
        \begin{itemize} 
            \item $\mathsf{scm}^{\mathsf{user}}_{i}$ := $\mathsf{comm}_{\mathsf{blind}_{i}^{\mathsf{state}}}$($sk$, $H$, $\mathsf{ctr}_{i}$, $\mathsf{bal}_{i}$, $e_{i}$, $\mathsf{scm}^{\mathsf{user}}_{i-1}$, $\mathsf{ccm}_{i}$)
            \item $\mathsf{scm}^{\mathsf{user}}_{i+1}$ = $\mathsf{comm}_{\mathsf{blind}^{\mathsf{state}}_{i+1}}$($sk$, $H$, $\mathsf{ctr}_{i}$, $\mathsf{bal}_{i}$, $e$, $\mathsf{scm}^{\mathsf{user}}_{i}$, $c$)
            \item $\mathsf{Ver}(pk_{\mathsf{CB}}, \mathsf{scm}^{\mathsf{user}}_{i}, \sigma^{\mathsf{user}}_{i}) = \mathsf{True}$
        \end{itemize}
    \end{itemize}
  \end{multicols}
  \end{mdframed}
\end{minipage}
    \caption{Zero-knowledge proofs for state recovery and synchronization.}
    \label{fig:zkp_other}
\end{figure*}

In this section, we discuss the details of the state recovery and synchronization functions. 

\para{State Recovery}
If a user cannot get a signature for one of their states through a regular signature request because of a double spend dependency, the user can invoke the state recovery mechanism instead. Although the double spend state remains unsigned, state recovery allows a user to get their own state signed by revealing details of the affected payment and identifying themselves towards the central bank. 

To invoke state recovery, the user identifies their own payment completion state $\mathsf{scm}^{\mathsf{user}}_{i}$ that completes the counterfeit payment.  For state recovery, the user generates the related history $\mathsf{hist}_{\mathsf{rel}}$, similarly to a payment creation, to prove they executed all state transitions correctly. In addition, the user needs to disclose their identifier $\mathsf{id}^{\mathsf{user}}$ and the received payment value $v^{\mathsf{pm}}$ to the central bank. The user generates $\mathsf{zkp}^{\mathsf{recovery}}$ (see \cref{fig:zkp_other}) to prove that these values were computed correctly and belong to $\mathsf{scm}$. To show that the user's previous state $\mathsf{scm}^{\mathsf{user}}_{i-1}$ is correct and unaffected by the double spend, $\mathsf{zkp}^{\mathsf{recovery}}$ also proves that $\mathsf{scm}^{\mathsf{user}}_{i-1}$ has a signature. The user sends $\mathsf{zkp}^{\mathsf{recovery}}$, $\mathsf{hist}_{\mathsf{rel}}$, $\mathsf{id}^{\mathsf{user}}$, and $v^{\mathsf{pm}}$ to the central bank. 

The central bank evaluates the state recovery request by first verifying $\mathsf{zkp}^{\mathsf{recovery}}$ using $\mathsf{pcm}$ in $\mathsf{hist}_{\mathsf{rel}}[\mathsf{scm}^{\mathsf{user}}_{i}]$ together with $\mathsf{id}^{\mathsf{user}}$ and $v^{\mathsf{pm}}$. This ensures that $v^{\mathsf{pm}}$ and the user identity $\mathsf{id}^{\mathsf{user}}$ belong to the owner of $\mathsf{scm}^{\mathsf{user}}_{i}$. If this succeeds, the central bank continues to verify the related history and adds entries to the central bank ledger $L$ for any states in the history that are not yet recorded. Finally, the central bank generates a signature $\sigma_{i}$ over $\mathsf{scm}^{\mathsf{user}}_{i}$, adds a corresponding ledger entry, and sends $\sigma_{i}$ to the user.

The user adds the $\sigma$ to $\mathsf{hist}_{\mathsf{ext}}[\mathsf{scm}]$ to conclude the state recovery process.

\para{Synchronization}
To update the epoch, the user first obtains a challenge $c$ and the current epoch $e$ from the central bank and creates a state transition to update the epoch. Similar to the enrollment process, the user includes $c$ in place of the counterparty commitment. Next, the user computes the new state commitment $\mathsf{scm}_{i+1}^{\mathsf{user}}$ and generates $\mathsf{zkp}^{\mathsf{sync}}$ (see \cref{fig:zkp_other}) to show that they updated their state correctly, and that their previous user state is signed. 
The user then sends $\mathsf{zkp}^{\mathsf{sync}}$, $e$, and $c$, $\mathsf{scm}_{i+1}^{\mathsf{user}}$ to the central bank to request a signature. 

The central bank checks for existing entries in $L$ with $\mathsf{scm}_{i+1}^{\mathsf{user}}$ and a signature $\sigma_{i+1}^{\mathsf{user}}$ and returns $\sigma_{i+1}^{\mathsf{user}}$ if found. Otherwise, the central bank checks that $e$ and $c$ match the expected values and verifies $\mathsf{zkp}^{\mathsf{sync}}$. If this succeeds, it generates a signature $\sigma_{i+1}^{\mathsf{user}}$, updates its ledger $L$, and returns $\sigma_{i+1}^{\mathsf{user}}$ to the user. %
The user checks if $\sigma_{i+1}^{\mathsf{user}}$ is a valid signature and adds it to $\mathsf{hist}_{\mathsf{ext}}[\mathsf{scm}_{i+1}^{\mathsf{user}}]$ to conclude the synchronization process.

\subsection{Offline session unlinkability towards other users without synchronization}
\label{subsec:unlinkability-wo-sync}
We first describe the edge case in which unlinkability of offline sessions towards other users does not hold. Next, we explain how and why synchronization helps mitigate this issue towards other users, but not towards other users who collude with the central bank. Lastly, we describe a mitigation strategy that ensures offline session unlinkability towards other users, even when colluding with the central bank, and does not require synchronization.

\para{Problem Description}
The states of a user are only unlinkable across offline sessions in one particular instance. Consider the case of some user Alice who just made an offline payment to Bob. Alice reconnects, gets a signature over her most recent payment creation state, and goes offline again.
Alice then completes an offline payment from Carol. Since Carol's payment state is unsigned, Alice cannot yet generate a valid dependency ZKP for her new payment completion state even though this is her first payment since reconnecting.

Alice stays offline and does not reconnect. At some point, Alice makes a payment to Dora. The related history for this payment includes Carol's unsigned payment state and Alice's corresponding payment completion state. In addition, since Alice's payment completion state does not yet have a dependency ZKP, Alice must include the state commitment and signature from her payment to Bob.

Therefore, if Bob ever becomes a direct or indirect recipient of Alice's payment to Dora, Bob can recognize and link the state commitment across the first offline session, before Alice's reconnected, and the second offline session, after Alice's reconnected.

\para{Partial mitigation through synchronization}
A simple mitigation strategy that can be adopted without modifying the protocol is to let users synchronize when they reconnect. This way, Alice's last state after reconnecting is a synchronization state. Since Alice did not share this synchronization state commitment  with any user of the first offline session, none of the (direct or indirect) recipients of her payment creation in the second offline session can link her state to another state from the first session anymore. 

However, if we consider cases in which the central bank can link multiple signature requests or ledger queries from the same user \textit{and} the central bank colludes with users, this mitigation strategy is not sufficient. In this case, the central bank can link Alice's synchronization state commitment to the previous state commitments of the user. If the central bank shares Alice's new synchronization state commitment with any of the recipient's of Alice's payments in the second session, e.g., Dora or Bob, they can again recognize Alice's state commitment and disclose her new state commitments from the second offline session to the central bank.

\para{Full mitigation through independent dependency ZKPs}
Finally, we discuss a full mitigation of the aforementioned leakage, which ensures offline session unlinkability towards other users, even if they collude with the central bank.

For this, we slightly modify the payment completion protocol: Instead of a single dependency ZKP, users create a separate dependency ZKP for each dependency state. This allows the user to hide the link to signed dependency states from recipients, even if another dependency is not yet signed.

This modification slightly increases the size of payment completion elements in the related history and a minor increases execution times. When users reconnect or complete a payment, they may require additional time to generate the separate dependency ZKPs. 
Before accepting a payment, users may also have to verify an additional dependency ZKP for payment completion states. 
Similarly, the central bank's processing time for payment completion states would increase, since the central bank needs to verify the two dependency ZKPs separately. However, should the increased total verifier runtime for users and the central bank become a concern, techniques such as aggregate ZKPs could be used to reduce overhead.

\subsection{Pseudocode}
\label{sec:sel_pseudocode}

In the following we provide pseudocode for all functions used in our protocol. \Cref{fig:pseudocode} shows the functions needed for reconnect and state recovery. \Cref{fig:pseudo_pay} shows the main functions needed to execute an offline payment, which make use of some helper functions in \cref{fig:pseudo_helper}. \Cref{fig:pseudo_pay_proof} shows the functions needed to generate the payment proofs. \Cref{fig:pseudo_on} shows the enrollment and synchronization functions. \Cref{fig:pseudo_CB} shows all central bank functions. \Cref{fig:pseudo_ver} shows the helper functions needed to verify states and state transitions.

\newenvironment{func}[1]
    {%
        \begin{subfigure}{0.45\textwidth}
                \underline{#1}
                \begin{enumerate}
    }{
                \end{enumerate}
                \vspace{1em}
            \end{subfigure}
    }

\begin{figure*} [htb]
\noindent 
\begin{minipage}[t]{\linewidth} 
\begin{mdframed}
\small
\begin{multicols}{2}

\begin{func}{$U.\mathsf{Reconnect}$($\mathsf{scm}_{i}^{\mathsf{user}}$)}  
    \item (flag, $\mathsf{scm}^{\mathsf{recovery}}$) := (false, $\bot$)
    \item If $U.\mathsf{QuerySignature}(\mathsf{scm}_{i}^{\mathsf{user}})$, return (false, $\bot$).
    \item If $\mathsf{hist}_{\mathsf{ext}}[\mathsf{scm}_{i}^{\mathsf{user}}]$ does not contain $\mathsf{zkp}^{\mathsf{dep}}$:
    \begin{itemize}
        \item (flag, $\mathsf{scm}^{\mathsf{recovery}}$) = $U.\mathsf{Reconnect}(\mathsf{scm}_{i-1}^{\mathsf{user}})$
        \item If flag is true, return (flag, $\mathsf{scm}^{\mathsf{recovery}}$) %
        \item If $\mathsf{scm}_{i}^{\mathsf{user}}$ corresponds to a payment completion state:
        \begin{itemize}
            \item (flag, $\mathsf{scm}^{\mathsf{recovery}}$) = $U.\mathsf{Reconnect}(\mathsf{ccm}_{i})$
            \item If flag is true: 
            \begin{itemize}
                \item If $\mathsf{scm}_{i}^{\mathsf{user}}$ is owned by the user, return (flag, $\mathsf{scm}_{i}^{\mathsf{user}}$). %
                \item Else, return (flag, $\mathsf{scm}^{\mathsf{recovery}}$).
            \end{itemize}
        \end{itemize}
    \end{itemize}
        \item $m^{\mathsf{sig\_request}} := U.\mathsf{CreateSigRequest}(\mathsf{scm}_{i}^{\mathsf{user}})$
        \item Send $m^{\mathsf{sig\_request}}$ to the central bank and let the response be $\sigma$.
        \item If $U.\mathsf{VerifyResponse}(\mathsf{scm}_{i}^{\mathsf{user}}, \sigma)$ returns true, add $\sigma$ to $\mathsf{hist}_{\mathsf{ext}}[\mathsf{scm}_{i}^{\mathsf{user}}]$ and return (false, $\bot$). 
        \item Else, return (true, $\bot$). 
\end{func}

\begin{func}{$U.\mathsf{QuerySignature}(\mathsf{scm})$}
    \item If $\mathsf{hist}_{\mathsf{ext}}[\mathsf{scm}].\sigma \ne \bot$, return true.
    \item Query ledger for a signature $\sigma$ over $\mathsf{scm}$. 
    \item If the ledger contains a valid $\sigma$ for $\mathsf{scm}$, add it to $\mathsf{hist}_{\mathsf{ext}}[\mathsf{scm}]$ and return true.
    \item Else, return false.
\end{func}

\begin{func}{$U.\mathsf{VerifyResponse}(\mathsf{scm}, \sigma)$}
    \item If $\sigma$ is not a valid central bank signature for $\mathsf{scm}$, return false. 
    \item Else, add $\sigma$ to $\mathsf{hist}_{\mathsf{ext}}[\mathsf{scm}]$ and return true. 
\end{func}

\vfill\null \columnbreak
\begin{func}{$U.\mathsf{CreateSigRequest}(\mathsf{scm})$}
    \item If $\mathsf{hist}_{\mathsf{ext}}[\mathsf{scm}]$ does not contain $\mathsf{zkp}^{\mathsf{dep}}$
    \begin{itemize}
        \item For every dependency of $\mathsf{scm}$ with state commitment $\mathsf{scm'}$
        \begin{itemize}
            \item If $\mathsf{hist}_{\mathsf{ext}}[\mathsf{scm'}].\sigma = \bot$, return $\bot$
        \end{itemize}
        \item Compute $\mathsf{zkp}^{\mathsf{dep}}$ and store the result in $\mathsf{hist}_{\mathsf{ext}}[\mathsf{scm}]$
    \end{itemize}
    \item If $\mathsf{scm}$ corresponds to a payment creation state, return $(\mathsf{PaymentCreation}$, $\mathsf{scm}$, $\mathsf{dcm}$, $\mathsf{sn}$, $\mathsf{ds}$, $\mathsf{zkp}^{\mathsf{state}}$, $\mathsf{zkp}^{\mathsf{dep}})$
    \item Else, return $\breaktuple{\mathsf{Payment Completion}$, $\mathsf{scm}$, $\mathsf{dcm}$, $\mathsf{pcm}$, $\mathsf{zkp}^{\mathsf{correct}}$, $\mathsf{zkp}^{\mathsf{dep}}$, $\mathsf{zkp}^{\mathsf{pm}}}$
\end{func}

\begin{func}{$U.\mathsf{StateRecovery}(\mathsf{scm})$}
    \item If $\mathsf{scm}$ is not owned by the user, return false.
    \item Obtain related history $\mathsf{hist}_{\mathsf{rel}}=U.\mathsf{GetRelatedHistory}(\mathsf{scm})$ 
    \item Generate $\mathsf{zkp}^{\mathsf{recovery}}$ using $\mathsf{hist}_{\mathsf{ext}}[\mathsf{scm}]$, $\mathsf{hist}_{\mathsf{int}}[\mathsf{scm}]$, and $\mathsf{hist}_{\mathsf{recovery}}[\mathsf{scm}]$  (See \cref{fig:zkp_other}).
    \item Send $(\mathsf{State Recovery}$, $\mathsf{scm}$, $\mathsf{id}^{\mathsf{user}}$, $v^{\mathsf{pm}}$, $\mathsf{zkp}^{\mathsf{recovery}}$, $\mathsf{hist}_{\mathsf{rel}})$ to the central bank.
    \item If the central bank returns $\sigma$, set $\mathsf{hist}_{\mathsf{ext}}[\mathsf{scm}]=\sigma$ and return true. Else, return false.
\end{func}
  \end{multicols}
  \end{mdframed}
\end{minipage}
    \caption{Pseudocode for user reconnect function with functions that reconnect calls.}
    \label{fig:pseudocode}
\end{figure*}

\begin{figure*} [htb]
\noindent 
\begin{minipage}[t]{\linewidth} 
\begin{mdframed}
\small
\begin{multicols}{2}
\begin{func}{$U.\mathsf{RequestPayment}(v^{\mathsf{pm}})$}
    \item Check that the wallet is not expired 
    \item Let $\mathsf{scm}^{\mathsf{rec}}_{k}$ be the current state commitment of the user
    \item Choose a request blind $\mathsf{blind}^\mathsf{req}_{k}$
    \item Compute the request commitment by committing to the current state commitment $\mathsf{rcm}^{\mathsf{rec}}_{k}$ := $\mathsf{comm}_{\mathsf{blind}^\mathsf{req}_{k}}$($\mathsf{rcm}^{\mathsf{rec}}_{k}$)
    \item  Store $\mathsf{blind}^\mathsf{req}_{k}$ in the internal history $\mathsf{hist}_{\mathsf{int}}[\mathsf{scm}^{\mathsf{rec}}_{k}]$
    \item Send $\mathsf{rcm}^{\mathsf{rec}}_{k}$ and the requested value $v^{\mathsf{pm}}$ to the recipient. 
\end{func}
\begin{func}{$U.\mathsf{CreatePayment}(\mathsf{rcm}^{\mathsf{rec}}_{\mathsf{k}}, v^{\mathsf{pm}})$}
    \item Check that the wallet is not expired
    \item Choose a random $\mathsf{blind}^{\mathsf{state}}_{l+1}$
    \item Create $\mathsf{state}^{\mathsf{sen}}_{l+1}$ as a state transition from the most recent user state $\mathsf{state}^{\mathsf{sen}}_{l}$
    \item Compute $\mathsf{scm}^{\mathsf{sen}}_{l+1}$:=$\mathsf{comm}_{\mathsf{blind}^{\mathsf{state}}_{l+1}}$($sk$, $H$, $\mathsf{ctr}_{l} + 1$, $\mathsf{bal}_{l} - v^{\mathsf{pm}}$, $e_{l}$, $\mathsf{scm}^{\mathsf{sen}}_{l}$, $\mathsf{rcm}^{\mathsf{rec}}_{\mathsf{k}}$)
    \item Store the openings of $\mathsf{scm}^{\mathsf{sen}}_{l+1}$ in $\mathsf{hist}_{\mathsf{int}}[\mathsf{scm}^{\mathsf{sen}}_{l+1}]$.
    \item Generate and update the external data for the new payment creation state: 
    \begin{itemize}
        \item Choose random $\mathsf{blind}_{l+1}^{\mathsf{dep}}$   
        \item $m_{l+1}^{\mathsf{state}}$ = $U.\mathsf{GenCreationStateProof}$($\mathsf{blind}_{l+1}^{\mathsf{dep}}, v^{\mathsf{pm}}$)
        \item Add $\mathsf{blind}_{l+1}^{\mathsf{dep}}, \mathsf{scm}^{\mathsf{sen}}_{l}$ and the values of $m_{l+1}^{\mathsf{state}}$ to $\mathsf{hist}_{\mathsf{ext}}[\mathsf{scm}^{\mathsf{sen}}_{l+1}]$ 
        \item $m_{l+1}^{\mathsf{dep}} = U.\mathsf{GenCreationDependencyProof}(\mathsf{hist}_{\mathsf{ext}})$ 
        \item Merge the values of $m_{l+1}^{\mathsf{dep}}$ into $\mathsf{hist}_{\mathsf{ext}}[\mathsf{scm}^{\mathsf{sen}}_{l+1}]$ 
    \end{itemize}
    \item Obtain related history $\mathsf{hist}_{\mathsf{rel}}=U.\mathsf{GetRelatedHistory}(\mathsf{scm}^{\mathsf{sen}}_{l+1})$ 
    \item Generate the payment proof message $m_{l+1}^{\mathsf{pm}} = {U.}\mathsf{GenPaymentProof}({ v^{\mathsf{pm}}})$ 
    \item Send $(\mathsf{hist}_{\mathsf{rel}}, m_{l+1}^{\mathsf{pm}})$ to the recipient 
\end{func}
\vfill\null \columnbreak

\begin{func}{$U.\mathsf{{AcceptPayment}}(\mathsf{hist}_{\mathsf{rel}}, m_{l+1}^{\mathsf{pm}})$}
    \item Parse $(\mathsf{scm}^{\mathsf{sen}}_{l+1}, v^{\mathsf{pm}}, e^{\mathsf{sen}}_{l}, \mathsf{blind}_{l+1}^{\mathsf{pm}}, \mathsf{zkp}^{pm})$ from $m_{l+1}^{\mathsf{pm}}$
    \item Parse $(\mathsf{sn}_{l+1}, \mathsf{ds}_{l+1}, \mathsf{dcm}^{\mathsf{sen}}_{l+1}, \mathsf{zkp}^{\mathsf{state}}_{l+1})$ from $\mathsf{hist}_{\mathsf{rel}}[\mathsf{scm}^{\mathsf{sen}}_{l+1}]$ 
    \item Let $\mathsf{scm}^{\mathsf{rec}}_{k}$ be the most recent state of the user
    \item Parse $\mathsf{blind}^\mathsf{req}_{k}$ from $\mathsf{hist}_{\mathsf{int}}[\mathsf{scm}^{\mathsf{rec}}_{k}]$
    \item Check that the sender's epoch $e^{\mathsf{sen}}_{l}$ and the user's epoch $e_{k}$ are recent enough 
    \item Verify the payment 
    \begin{enumerate}
        \item Verify the payment proof $\mathsf{zkp}^{\mathsf{pm}}$
        \begin{itemize}
            \item Compute $\mathsf{rcm}^{\mathsf{rec}}_{k} = \mathsf{comm}_{\mathsf{blind}^\mathsf{req}_{k}}(\mathsf{scm}^{\mathsf{rec}}_{k})$
            \item Compute $\mathsf{pcm}_{l+1}^{\mathsf{sen}}$ = $\mathsf{comm}_{\mathsf{blind}_{l+1}^{\mathsf{pm}}}$($v^{\mathsf{pm}}$, ${\mathsf{rcm}^{\mathsf{rec}}_{k}}$, $\mathsf{scm}^{\mathsf{sen}}_{l+1}$, $e^{\mathsf{sen}}_{l})$
            \item Verify $\mathsf{zkp}^{\mathsf{pm}}$ for public input $\mathsf{pcm}_{l+1}^{\mathsf{sen}}$
        \end{itemize}
        \item Verify the validity of the new sender state $\mathsf{VerifyOfflineCreation}(\mathsf{hist}_{\mathsf{rel}}, \mathsf{scm}^{\mathsf{sen}}_{l+1}$)
        \item If any check fails, reject the payment. Else, accept the payment.
   \end{enumerate}
\end{func}
\begin{func}{$U.\mathsf{{CompletePayment}}(\mathsf{hist}_{\mathsf{rel}}, m^{\mathsf{pm}})$}
    {\item Parse $(\mathsf{scm}^{\mathsf{sen}}_{l+1}, v^{\mathsf{pm}}, e^{\mathsf{sen}}_{l}, \mathsf{blind}_{l+1}^{\mathsf{pm}}, \mathsf{zkp}^{pm})$ from $m^{\mathsf{pm}}$
    \item Parse $\mathsf{blind}^\mathsf{req}_{k}$ from $\mathsf{hist}_{\mathsf{int}}[\mathsf{scm}^{\mathsf{rec}}_{k}]$
    \item Compute $\mathsf{rcm}^{\mathsf{rec}}_{k} = \mathsf{comm}_{\mathsf{blind}^\mathsf{req}_{k}}(\mathsf{scm}^{\mathsf{rec}}_{k})$
    \item Compute $\mathsf{pcm} = \mathsf{comm}_{\mathsf{blind}_{l+1}^{\mathsf{pm}}}(v^{\mathsf{pm}}, \mathsf{rcm}^{\mathsf{rec}}_{k}, \mathsf{scm}^{\mathsf{sen}}_{l+1}, e^{\mathsf{sen}}_{l})$}
    \item Choose a random $\mathsf{blind}^{\mathsf{state}}_{k+1}$
    \item Update the user state by creating $\mathsf{state}^{\mathsf{rec}}_{k+1}$ as a state transition from $\mathsf{state}^{\mathsf{rec}}_{k}$
    \item Generate the new state commitment $\mathsf{scm}^{\mathsf{rec}}_{k+1}=\mathsf{comm}_{\mathsf{blind}^{\mathsf{state}}_{k+1}}(sk, H, {\mathsf{ctr}_{k}}, \mathsf{bal}_{k}+v^{\mathsf{pm}}, e_{k}, \mathsf{scm}^{\mathsf{rec}}_{k}, \mathsf{scm}^{\mathsf{sen}}_{l+1})$
    \item Store the openings of $\mathsf{scm}^{\mathsf{rec}}_{k+1}$ in $\mathsf{hist}_{\mathsf{int}}[\mathsf{scm}^{\mathsf{rec}}_{k+1}]$.
    \item Update $\mathsf{hist}_{\mathsf{ext}}$ by merging the related history $\mathsf{hist}_{\mathsf{rel}}$ into it and adding $\mathsf{pcm}, \mathsf{zkp}^{\mathsf{pm}}$ to $\mathsf{hist}_{\mathsf{ext}}[\mathsf{scm}^{\mathsf{rec}}_{k+1}]$.
    \item Add $\mathsf{blind}_{l+1}^{\mathsf{pm}}, v^{\mathsf{pm}}, e^{\mathsf{sen}}_{l}$ to $\mathsf{hist}_{\mathsf{recovery}}[\mathsf{scm}^{\mathsf{rec}}_{k+1}]$
    \item Add external data for the current payment completion state: 
    \begin{itemize}
        \item Choose random $\mathsf{blind}_{k+1}^{\mathsf{dep}}$   
        \item Set $m_{k+1}^{\mathsf{state}}$ = ${U.}\mathsf{GenCompletionStateProof}$ ($\mathsf{blind}_{k+1}^{\mathsf{dep}}$, $\mathsf{blind}_{l+1}^{\mathsf{pm}}$, $v^{\mathsf{pm}}$, $\mathsf{pcm}_{l+1}^{\mathsf{sen}}$, $e^{\mathsf{sen}}_{l}$)
        \item Add $({\mathsf{blind}_{k+1}^{\mathsf{dep}}}, {\mathsf{scm}^{\mathsf{rec}}_{k}}, \mathsf{scm}^{\mathsf{sen}}_{l+1})$ and the values of $m^{\mathsf{state}}$ to $\mathsf{hist}_{\mathsf{ext}}[\mathsf{scm}^{\mathsf{rec}}_{k+1}]$ 
        \item Set $m_{k+1}^{\mathsf{dep}}$ = $U.\mathsf{GenCompletionDependencyProof}$ ($\mathsf{state}^{\mathsf{rec}}_{k+1}$, $\mathsf{hist}_{\mathsf{ext}}$) 
        \item Add the values of $m_{k+1}^{\mathsf{dep}}$ to $\mathsf{hist}_{\mathsf{ext}}[\mathsf{scm}^{\mathsf{rec}}_{k+1}]$ 
    \end{itemize}
\end{func}
  \end{multicols}
  \end{mdframed}
\end{minipage}
    \caption{Pseudocode for offline payment functions.}
    \label{fig:pseudo_pay}
\end{figure*}

\begin{figure*} [htb]
\noindent 
\begin{minipage}[t]{\linewidth} 
\begin{mdframed}
\small
\begin{multicols}{2}
\begin{func}{${U.}\mathsf{GenCreationStateProof}\allowbreak({\mathsf{blind}_{l+1}^{\mathsf{dep}}, v^{\mathsf{pm}}})$}
    \item Compute dependency commitment $\mathsf{dcm}^{\mathsf{sen}}_{l+1} = \mathsf{comm}_{\mathsf{blind}_{l+1}^{\mathsf{dep}}}(\mathsf{scm}^{\mathsf{sen}}_{l})$
    \item Compute $\mathsf{sn}_{l+1}=\mathsf{PRF}^{\mathsf{sn}}_{sk}(\mathsf{ctr}_{l+1})$  
    \item Compute $\mathsf{ds}_{l+1}=\mathsf{id}^{\mathsf{sen}} + \mathsf{scm}^{\mathsf{sen}}_{l+1} \cdot \mathsf{PRF}^{\mathsf{ds}}_{sk}(\mathsf{ctr}_{l+1})$
    \item Generate $\mathsf{zkp}^{\mathsf{state}}$  to prove that the state transition was computed correctly (see $\mathsf{zkp}^{\mathsf{state}}_{\mathsf{create}}$ in \cref{fig:zkp_main})
    \item Return $m_{l+1}^{\mathsf{state}} = (\mathsf{sn}_{l+1}, \mathsf{ds}_{l+1}, \mathsf{scm}^{\mathsf{sen}}_{l+1}, \mathsf{dcm}^{\mathsf{sen}}_{l+1}, \mathsf{zkp}^{\mathsf{state}})$ 
\end{func}
\begin{func}{
    ${U.}\mathsf{GenCompletionStateProof}({\mathsf{blind}_{k+1}^{\mathsf{dep}}, \mathsf{blind}_{l+1}^{\mathsf{pm}}, v^{\mathsf{pm}}, \mathsf{pcm}, e^{\mathsf{sen}}_{l}})$}
    \item Compute dependency commitment $\mathsf{dcm}^{\mathsf{rec}}_{k+1} = \mathsf{comm}_{\mathsf{blind}_{k+1}^{\mathsf{dep}}}(\mathsf{scm}^{\mathsf{rec}}_{k}, \mathsf{scm}^{\mathsf{sen}}_{l+1})$
    \item Parse  $\mathsf{blind}^\mathsf{req}_{k}$ from $\mathsf{hist}_{\mathsf{int}}[\mathsf{scm}^{\mathsf{rec}}_{k}]$
    \item Generate $\mathsf{zkp}^{\mathsf{state}}$  to prove that the state transition was computed correctly (see $\mathsf{zkp}^{\mathsf{state}}_{\mathsf{create}}$ in \cref{fig:zkp_main})
    \item Return $m_{k+1}^\mathsf{state}=(\mathsf{scm}^{\mathsf{rec}}_{k+1}, \mathsf{dcm}^{\mathsf{rec}}_{k+1}, \mathsf{pcm}^{\mathsf{sen}}_{l+1}, \mathsf{zkp}^{\mathsf{state}})$  
\end{func}
\begin{func}{${ U.}\mathsf{GenPaymentProof}(v^{\mathsf{pm}})$}
    \item Choose random $\mathsf{blind}^{pm}_{l+1}$ to generate $\mathsf{pcm}^{\mathsf{sen}}_{l+1} = \mathsf{comm}_{\mathsf{blind}_{l+1}^{\mathsf{pm}}}(v^{\mathsf{pm}}, \mathsf{ccm}_{l+1}, \mathsf{scm}^{\mathsf{sen}}_{l+1}, e_{l})$.
    \item Generate $\mathsf{zkp}^{\mathsf{pm}}$ to allow the recipient to complete the payment (see $\mathsf{zkp}^{\mathsf{pm}}$ in \cref{fig:zkp_main})
    \item Return $m^{\mathsf{pm}}_{l+1}=(\mathsf{scm}^{\mathsf{sen}}_{l+1}, v^{\mathsf{pm}}, e_{l}, \mathsf{blind}^{\mathsf{pm}}_{l+1}, \mathsf{zkp}^{\mathsf{pm}})$ 
\end{func}
\vfill\null \columnbreak
\begin{func}{$U.\mathsf{GenCreationDependencyProof}(\mathsf{scm}^{\mathsf{sen}}_{l}, \mathsf{hist}_{\mathsf{ext}})$}
    \item Get the dependency blind $\mathsf{blind}^{dep}_{l}$ and the previous state commitment $\mathsf{scm}^{\mathsf{sen}}_{l-1}$ from $\mathsf{hist}_{\mathsf{ext}}[\mathsf{scm}^{\mathsf{sen}}_{l}]$
    \item If $\mathsf{scm}^{\mathsf{sen}}_{l-1}$ does not yet have a signature $\sigma^{\mathsf{sen}}_{l-1}$, return $\bot$.
    \item Compute the dependency commitment $\mathsf{dcm}^{\mathsf{sen}}_{l} = \mathsf{comm}_{\mathsf{blind}_{l}^{\mathsf{dep}}}(\mathsf{scm}_{l-1}^{\mathsf{sen}})$
    \item Generate $\mathsf{zkp}^{\mathsf{dep}}$  to prove that all dependencies of the state are signed (see $\mathsf{zkp}^{\mathsf{dep}}_{\mathsf{create}}$ in \cref{fig:zkp_main})
    \item Return $m^{\mathsf{dep}}_{l} = (\mathsf{zkp}^{\mathsf{dep}}, \mathsf{dcm}^{\mathsf{sen}}_{l})$ 
\end{func}
\begin{func}{$U.\mathsf{GenCompletionDependencyProof}({\mathsf{scm}^{\mathsf{rec}}_{k}, \mathsf{hist}_{\mathsf{ext}}})$}
   \item Get the blind $\mathsf{blind}^{\mathsf{dep}}_{k}$, the previous state commitment $\mathsf{scm}^{\mathsf{rec}}_{k-1}$, and the counterparty state commitment $\mathsf{ccm}_{k}$ from $\mathsf{hist}_{\mathsf{ext}}[\mathsf{scm}^{\mathsf{rec}}_{k}]$
    \item If either $\mathsf{scm}^{\mathsf{rec}}_{k-1}$ or $\mathsf{ccm}_{k}$ do not yet have a signature $\sigma^{\mathsf{rec}}_{k-1}$ or $\sigma^{\mathsf{cp}}_{k}$, return $\bot$.
    \item Compute dependency commitment $\mathsf{dcm}^{\mathsf{rec}}_{k} = \mathsf{comm}_{\mathsf{blind}_{k}^{\mathsf{dep}}}(\mathsf{scm}^{\mathsf{rec}}_{k-1}, \mathsf{ccm}_{k})$
    \item Generate {$\mathsf{zkp}^{\mathsf{dep}}$} to prove that all dependencies of the state are signed (See $\mathsf{zkp}_{\mathsf{comp}}^{\mathsf{dep}}$ in \cref{fig:zkp_main})
    \item Return $m^{\mathsf{dep}}_{k} = (\mathsf{zkp}^{\mathsf{dep}}, \mathsf{dcm}^{\mathsf{rec}}_{k})$ 
\end{func}
  \end{multicols}
  \end{mdframed}
\end{minipage}
    \caption{Pseudocode for proof creation.}
    \label{fig:pseudo_pay_proof}
\end{figure*}

\begin{figure*} [htb]
\noindent 
\begin{minipage}[t]{\linewidth} 
\begin{mdframed}
\small
\begin{multicols}{2}

\begin{func}{$U.\mathsf{GetRelatedHistory}(\mathsf{scm}_{i}^{\mathsf{user}})$}
    \item $\mathsf{hist}_{\mathsf{rel}} = \{U.\mathsf{GetElement}(\mathsf{scm}_{i}^{\mathsf{user}})\}$
    \item If $\mathsf{hist}_{\mathsf{rel}}$ does not contain $\sigma_{i}$ or $\mathsf{zkp}^{\mathsf{dep}}$:
    \begin{itemize}
        \item Set $\mathsf{hist}_{\mathsf{rel}} = \mathsf{hist}_{\mathsf{rel}} \cup U.\mathsf{GetRelatedHistory}(\mathsf{scm}_{i-1}^{\mathsf{user}})$
        \item If $\mathsf{scm}_{i}^{\mathsf{user}}$ corresponds to a payment completion state, set $\mathsf{hist}_{\mathsf{rel}} = \mathsf{hist}_{\mathsf{rel}} \cup U.\mathsf{GetRelatedHistory}(\mathsf{ccm}_{i})$
    \end{itemize}
    \item Return $\mathsf{hist}_{\mathsf{rel}}$
\end{func}
\vfill\null \columnbreak

\begin{func}{$U.\mathsf{GetElement}(\mathsf{scm}_{i}^{\mathsf{user}})$}
    \item If $\mathsf{hist}_{\mathsf{ext}}[\mathsf{scm}_{i}^{\mathsf{user}}]$ contains $\sigma_{i}$, return $(\mathsf{scm}_{i}^{\mathsf{user}}, \sigma_{i})$    
    \item If $\mathsf{scm}_{i}^{\mathsf{user}}$ corresponds to a payment creation state
    \begin{itemize}
        \item If $\mathsf{hist}_{\mathsf{ext}}[\mathsf{scm}_{i}^{\mathsf{user}}]$ contains $\mathsf{zkp}^{\mathsf{dep}}$, return $(\mathsf{sn}_{i}, \mathsf{ds}_{i}, \mathsf{scm}_{i}^{\mathsf{user}}, \mathsf{dcm}_{i}^{\mathsf{user}}, \mathsf{zkp}^{\mathsf{correct}}, \mathsf{zkp}^{\mathsf{dep}})$
        \item Else, return $(\mathsf{sn}_{i}, \mathsf{ds}_{i}, \mathsf{scm}_{i}^{\mathsf{user}}, \mathsf{zkp}^{\mathsf{correct}}, \mathsf{blind}_{i}^{\mathsf{dep}}, \mathsf{scm}_{i-1}^{\mathsf{user}})$
    \end{itemize}
    \item If $\mathsf{scm}_{i}^{\mathsf{user}}$ corresponds to a payment completion state 
    \begin{itemize}
         \item if $\mathsf{hist}_{\mathsf{ext}}[\mathsf{scm}_{i}^{\mathsf{user}}]$ contains $\mathsf{zkp}^{\mathsf{dep}}$, return $(\mathsf{scm}_{i}^{\mathsf{user}}, \mathsf{dcm}_{i}^{\mathsf{user}}, \mathsf{pcm}_{i}, \mathsf{zkp}^{\mathsf{correct}}, \mathsf{zkp}^{\mathsf{\mathsf{pm}}}, \mathsf{zkp}^{\mathsf{dep}})$ 
         \item Else, return $\breaktuple{\mathsf{scm}_{i}^{\mathsf{user}}, \mathsf{pcm}_{i}, \mathsf{zkp}^{\mathsf{correct}}, \mathsf{zkp}^{\mathsf{\mathsf{pm}}}, \mathsf{blind}_{i}^{\mathsf{dep}}, \mathsf{scm}_{i-1}^{\mathsf{user}}, \mathsf{ccm}_{i}}$ 
    \end{itemize}
    \item Else, return $()$ 
\end{func}
  \end{multicols}
  \end{mdframed}
\end{minipage}
    \caption{Pseudocode for generating the related history for payment creations}
    \label{fig:pseudo_helper}
\end{figure*}

\begin{figure*} [htb]
\noindent 
\begin{minipage}[t]{\linewidth} 
\begin{mdframed}
\small
\begin{multicols}{2}
\begin{func}{$U.\mathsf{Enroll}()$} 
    \item Obtain the current epoch $e$ and a challenge $c$ from the central bank
    \item Choose random $sk$ and $\mathsf{blind}^{\mathsf{state}}_{i}$
    \item Compute ${\mathsf{id}}^{\mathsf{user}}$ as $\mathsf{PRF}_{sk}^{\mathsf{id}}(0)$
    \item Compute initial state commitment $\mathsf{scm}^{\mathsf{user}}_{i} =\mathsf{comm}_{\mathsf{blind}_{i}^{\mathsf{state}}}(sk, H, 0, e, 0, c)$
    \item Generate $\mathsf{zkp}^{\mathsf{enroll}}$ (see $\mathsf{zkp}^{\mathsf{enroll}}$ in \cref{fig:zkp_main})
    \item Store the openings of $\mathsf{scm}^{\mathsf{user}}_{i}$ in $\mathsf{hist}_{\mathsf{int}}[\mathsf{scm}^{\mathsf{user}}_{i}]$
    \item Send $m^{\mathsf{enroll}}_{i}= (\mathsf{zkp}^{\mathsf{enroll}}, \mathsf{id}^{\mathsf{user}}, \mathsf{scm}^{\mathsf{user}}_{i}, e, H, c)$ to the central bank to request enrollment.
    \item Let $\sigma^{\mathsf{user}}_{i}$ be the signature received from the central bank. 
    \item Store $\sigma^{\mathsf{user}}_{i}$ in $\mathsf{hist}_{\mathsf{ext}}[\mathsf{scm}^{\mathsf{user}}_{i}]$.
\end{func}
\vfill\null \columnbreak
\begin{func}{$U.\mathsf{Synchronize}$()}
    \item Obtain the current epoch $e$ and a challenge $c$ from the central bank
    \item Let $\mathsf{state}^{\mathsf{user}}_{i}$ be the most recent user state
    \item Choose a random {$\mathsf{blind}_{i+1}^{\mathsf{state}}$}
    \item Create a state transition from $\mathsf{state}^{\mathsf{user}}_{i}$ to $\mathsf{state}^{\mathsf{user}}_{i+1}$
    \item Compute $\mathsf{scm}^{\mathsf{user}}_{i+1} = \mathsf{comm}_{\mathsf{blind}_{i+1}^{\mathsf{state}}}(sk, H, \mathsf{ctr}_{i}, \mathsf{bal}_{i}, e, \mathsf{scm}^{\mathsf{user}}_{i}, c)$
    \item Generate $\mathsf{zkp}^{{\mathsf{sync}}}$ to prove that the state transition was computed correctly (see $\mathsf{zkp}^{\mathsf{sync}}$ in \cref{fig:zkp_other})
    \item Set $m^{\mathsf{sync}}_{i+1}=({\mathsf{scm}^{\mathsf{user}}_{i+1}}, e, c, \mathsf{zkp}^{{\mathsf{sync}}})$ 
    \item Store the openings of $\mathsf{scm}^{\mathsf{user}}_{i+1}$ in $\mathsf{hist}_{\mathsf{int}}[\mathsf{scm}^{\mathsf{user}}_{i+1}]$.
    \item Send $m^{\mathsf{sync}}_{i+1}$ to the central bank
    \item Let $\sigma^{\mathsf{user}}_{i+1}$ be the response of the central bank
    \item Store $\sigma^{\mathsf{user}}_{i+1}$ in $\mathsf{hist}_{\mathsf{ext}}[\mathsf{scm}^{\mathsf{user}}_{i+1}]$
\end{func}
  \end{multicols}
  \end{mdframed}
\end{minipage}
    \caption{Pseudocode for enroll and synchronization functions.}
    \label{fig:pseudo_on}
\end{figure*}

\begin{figure*} [htb]
\noindent 
\begin{minipage}[t]{\linewidth} 
\begin{mdframed}
\small
\begin{multicols}{2}
\begin{func}{$CB.\mathsf{CheckEnroll}(m_{0}^{\mathsf{enroll}})$}    
    \item Parse $m_{0}^{\mathsf{enroll}}$ as $(\mathsf{zkp}^{\mathsf{enroll}}, \mathsf{id}^{\mathsf{user}}, \mathsf{scm}^{\mathsf{user}}_{0}, e, H, c)$ and ensure that $H$, $e$, and $c$ match the expected values.
    \item Verify $\mathsf{zkp}^{\mathsf{enroll}}$ for the public inputs provided.
    \item If $\mathsf{id}^{\mathsf{user}}$ is not yet in the user registry $R$, add $\mathsf{id}^{\mathsf{user}}$ to $R$. Else, return false.
    \item Generate $\sigma^{\mathsf{user}}_{0} = \mathsf{Sign}(sk_{\mathsf{CB}}, \mathsf{scm}^{\mathsf{user}}_{0})$, add $(\mathsf{scm}^{\mathsf{user}}_{0}, \sigma^{\mathsf{user}}_{0})$ to the public ledger $L$, and send $\sigma^{\mathsf{user}}_{0}$ to the user.
\end{func}
\begin{func}{$CB.{\mathsf{ProcessSynchronizationRequest}}(m_{i}^{{\mathsf{sync}}})$}
    \item Parse $\breaktuple{\mathsf{scm}^{\mathsf{user}}_{i}, e, c, \mathsf{zkp}^{\mathsf{sync}}}$ from $m_{i}^{\mathsf{sync}}$.
    \item Check whether $e$ and $c$ match the expected values.     
    \item Verify $\mathsf{zkp}^{\mathsf{sync}}$ for public inputs
    \item If none of the entries in $L$ has the same state commitment $\mathsf{scm}^{\mathsf{user}}_{i}$, add $(\mathsf{scm}^{\mathsf{user}}_{i}, \bot)$ to $L$. Else if the entry has a signature $\sigma^{\mathsf{user}}_{i}$, return $\sigma^{\mathsf{user}}_{i}$. 
    \item Compute $\sigma^{\mathsf{user}}_{i} = \mathsf{Sign}(sk_{CB}, \mathsf{scm}^{\mathsf{user}}_{i})$ over $\mathsf{scm}^{\mathsf{user}}_{i}$, update the corresponding entry in $L$ and return $\sigma^{\mathsf{user}}_{i}$.
\end{func}
\begin{func}{$CB.\mathsf{IdentifyDoubleSpenders}(L)$}
    \item dsList = $\{\}$
    \item For any entries in $L$ with the same value $\mathsf{sn}$ and different state commitment $\mathsf{scm}\ne \mathsf{scm'}$, reconstruct the identity $\mathsf{id}$ of the double spender:
    \begin{itemize}
        \item Let $\mathsf{ds}, \mathsf{ds}'$ be the corresponding double spending tags with serial number $\mathsf{sn} := \mathsf{PRF}^{\mathsf{sn}}_{sk}(\mathsf{ctr})$
        \item Solve double spending equations $\mathsf{ds}'= \mathsf{id} + \mathsf{scm'} \cdot \mathsf{PRF}^{\mathsf{ds}}_{sk}(\mathsf{ctr})$  and $\mathsf{ds} = \mathsf{id} + \mathsf{scm} \cdot \mathsf{PRF}^{\mathsf{ds}}_{sk}(\mathsf{ctr})$ for $\mathsf{id}$.
        \item Add $\mathsf{id}$ to dsList
    \end{itemize}
    \item Return dsList
\end{func}
\vfill\null \columnbreak
\begin{func}{$CB.\mathsf{ProcessSignatureRequest}(m_{i}^{\mathsf{sig\_request}})$}
    \item If $\mathsf{VerifyState}(m_{i}^{\mathsf{sig\_request}})$ returns false, return false
    \item If the request is for a payment creation:
    \begin{itemize}
        \item Parse $(\mathsf{sn}_{i}, \mathsf{ds}_{i}, \mathsf{scm}^{\mathsf{user}}_{i}, \mathsf{dcm}^{\mathsf{user}}_{i}, \mathsf{zkp}^{\mathsf{state}})$ from $m_{i}^{\mathsf{sig\_request}}$ 
        \item If none of the entries in $L$ has the same state commitment $\mathsf{scm}^{\mathsf{user}}_{i}$, add $(\mathsf{sn}_{i}, \mathsf{ds}_{i}, \mathsf{scm}^{\mathsf{user}}_{i}, \bot)$ to $L$. Else if the entry has a signature $\sigma^{\mathsf{user}}_{i}$, return $\sigma^{\mathsf{user}}_{i}$. 
        \item If there is more than one entry in $L$ with $\mathsf{sn}$, return $\bot$ %
    \end{itemize}
    \item If the request is for a payment completion:
    \begin{itemize}
        \item Parse $(\mathsf{scm}^{\mathsf{user}}_{i}, \mathsf{dcm}^{\mathsf{user}}_{i}, \mathsf{pcm}, \mathsf{zkp}^{\mathsf{state}}, \mathsf{zkp}^{\mathsf{dep}}, \mathsf{zkp}^{\mathsf{pm}})$ from $m_{i}^{\mathsf{sig\_request}}$ 
        \item If none of the entries in $L$ has the same state commitment $\mathsf{scm}^{\mathsf{user}}_{i}$, add $(\mathsf{scm}^{\mathsf{user}}_{i}, \bot)$ to $L$. Else if an entry for $\mathsf{scm}^{\mathsf{user}}_{i}$ has a signature $\sigma^{\mathsf{user}}_{i}$, return $\sigma^{\mathsf{user}}_{i}$. 
    \end{itemize}
    \item Compute $\sigma^{\mathsf{user}}_{i} = \mathsf{Sign}(sk_{CB}, \mathsf{scm}^{\mathsf{user}}_{i})$, update the corresponding entry in $L$ and return $\sigma^{\mathsf{user}}_{i}$
\end{func}
\begin{func}{$CB.{\mathsf{ProcessStateRecoveryRequest}}(m_{i}^{\mathsf{recovery}})$}
    \item Parse $m_{i}^{\mathsf{recovery}}$ as $(\mathsf{scm}^{\mathsf{user}}_{i}, \mathsf{id}^{\mathsf{user}}, v^{\mathsf{pm}}, \mathsf{zkp}^\mathsf{recovery}, \mathsf{hist}_{\mathsf{rel}})$
    \item Parse $\mathsf{pcm}$ from $\mathsf{hist}_{\mathsf{rel}}[\mathsf{scm}]$ 
    \item Verify $\mathsf{zkp}^{\mathsf{recovery}}$ for public input $\mathsf{pcm}, \mathsf{scm}^{\mathsf{user}}_{i}, \mathsf{id}^{\mathsf{user}}, pk_{\mathsf{CB}}, v^{\mathsf{pm}}$
    \item If $\mathsf{VerifyOfflineCompletion}(\mathsf{hist}_{\mathsf{rel}}, \mathsf{scm}^{\mathsf{user}}_{i})$ returns false, return $\bot$.
    \item For every state commitment $\mathsf{scm'}$ in $\mathsf{hist}_{\mathsf{rel}}$:
    \begin{itemize}
        \item If $\mathsf{scm'}$ corresponds to a payment creation: 
        \begin{itemize}
            \item If none of the entries in $L$ has the same state commitment $\mathsf{scm'}$, use $\mathsf{hist}_{\mathsf{rel}}[\mathsf{scm'}]$ to add $(\mathsf{sn}, \mathsf{ds}, \mathsf{scm'}, \bot)$ to $L$ 
        \end{itemize}
        \item If $\mathsf{scm'}$ corresponds to a payment completion:
        \begin{itemize}
            \item If none of the entries in $L$ contains an entry with the same $\mathsf{scm'}$, add $(\mathsf{scm'}, \bot)$ to $L$ 
        \end{itemize}
    \end{itemize}
    \item Compute $\sigma^{\mathsf{user}}_{i} = \mathsf{Sign}(sk_{CB}, \mathsf{scm}^{\mathsf{user}}_{i})$, update the corresponding entry in $L$ and return $\sigma^{\mathsf{user}}_{i}$
\end{func}
  \end{multicols}
  \end{mdframed}
\end{minipage}
    \caption{Pseudocode for central bank functions.}
    \label{fig:pseudo_CB}
\end{figure*}

\begin{figure*} [htb]
\noindent 
\begin{minipage}[t]{\linewidth} 
\begin{mdframed}
\small
\begin{multicols}{2}
\begin{func}{$\mathsf{VerifyState}(m_{i}^{{\mathsf{state}}})$}
    \item Check that the state transition was performed correctly:
    \begin{itemize}
        \item If the request is for a payment creation:
        \begin{itemize}
            \item Parse $\mathsf{sn}_{i}, \mathsf{ds}_{i}, \mathsf{scm}^{\mathsf{user}}_{i}, \mathsf{dcm}^{\mathsf{user}}_{i}, \mathsf{zkp}^{\mathsf{state}}$ from $m_{i}^{\mathsf{state}}$
            \item Verify $\mathsf{zkp}^{\mathsf{state}}$ for public input $\mathsf{sn}_{i}, \mathsf{ds}_{i}, \mathsf{scm}^{\mathsf{user}}_{i}, \mathsf{dcm}^{\mathsf{user}}_{i}$
        \end{itemize}
        \item Else if the request is for a payment completion:
        \begin{itemize}
            \item Parse $\mathsf{scm}^{\mathsf{user}}_{i}, \mathsf{dcm}^{\mathsf{user}}_{i}, \mathsf{pcm}, \mathsf{zkp}^{\mathsf{pm}}, \mathsf{zkp}^{\mathsf{state}}$ from $m_{i}^{\mathsf{state}}$
            \item Verify $\mathsf{zkp}^{\mathsf{state}}$ for public input $\mathsf{pcm}$, $\mathsf{scm}^{\mathsf{user}}_{i}$, $\mathsf{dcm}^{\mathsf{user}}_{i}$, ${\Delta_{\mathsf{sync}}}$
            \item Verify $\mathsf{zkp}^{\mathsf{pm}}$ for public input $\mathsf{pcm}$
        \end{itemize}
    \end{itemize}
    \item Check that all dependencies are signed:
    \begin{itemize}
        \item Parse $\mathsf{zkp}^{\mathsf{dep}}, \mathsf{dcm}$ from $m_{i}^{\mathsf{state}}$ 
        \item Verify $\mathsf{zkp}^{\mathsf{dep}}$ for public input $\mathsf{dcm}^{\mathsf{user}}_{i}, pk_{\mathsf{CB}}$.
    \end{itemize}
    \item If any verification fails, return false. Else return true.
\end{func}
\vfill\null \columnbreak
\begin{func}{$\mathsf{VerifyOfflineCompletion}(\mathsf{hist}_{\mathsf{rel}}, \mathsf{scm}^{\mathsf{rec}}_{k})$}
    \item Let $m_{k}^{\mathsf{hist}}$ be the element with state commitment $\mathsf{scm}^{\mathsf{rec}}_{k}$ in $\mathsf{hist}_{\mathsf{rel}}$
    \item If $\mathsf{VerifyState}(m_{k}^{\mathsf{state}})$ is true, return true. %
    \item If $m_{k}^{\mathsf{hist}} = (\mathsf{scm}^{\mathsf{rec}}_{k}, \sigma^{\mathsf{rec}}_{k})$ and $\sigma^{\mathsf{rec}}_{k}$ is a valid central bank signature for $\mathsf{scm}^{\mathsf{rec}}_{k}$, return true. %
    \item Parse  $(\mathsf{scm}^{\mathsf{rec}}_{k}, \mathsf{dcm}^{\mathsf{rec}}_{k}, \mathsf{pcm}, \mathsf{zkp}^{\mathsf{pm}}, \mathsf{zkp}^{\mathsf{state}})$ from $m_{k}^{\mathsf{hist}}$ 
    \item Verify correctness of the payment completion  
    \begin{itemize}
        \item Verify $\mathsf{zkp}^{\mathsf{state}}$ for public input $\mathsf{pcm}, \mathsf{scm}^{\mathsf{rec}}_{k}, \mathsf{dcm}^{\mathsf{rec}}_{k}, \Delta_{\mathsf{sync}}$
        \item Verify $\mathsf{zkp}^{\mathsf{pm}}$ for public input $\mathsf{pcm}$
    \end{itemize}
    \item Parse $(\mathsf{blind}_{k}^{\mathsf{dep}}, \mathsf{scm}_{k-1}^{\mathsf{rec}}, \mathsf{ccm}_{k})$ from $m_{k}^{\mathsf{hist}}$
    \item Verify that $\mathsf{dcm}^{\mathsf{rec}}_{k}$ equals to $\mathsf{comm}_{\mathsf{blind}_{k}^{\mathsf{dep}}}(\mathsf{scm}^{\mathsf{rec}}_{k-1}, \mathsf{ccm}_{k})$
    \item If $\mathsf{scm}^{\mathsf{rec}}_{k-1}$ belongs to a payment creation state
    \begin{enumerate}
        \item If $\mathsf{VerifyOfflineCreation}(\mathsf{hist}_{\mathsf{rel}}, \mathsf{scm}^{\mathsf{rec}}_{k-1})$ and $\mathsf{VerifyOfflineCreation}(\mathsf{hist}_{\mathsf{rel}}, \mathsf{ccm}_{k})$ both return true, return true.
        \item Else, return false.
    \end{enumerate}
    \item Else
        \begin{enumerate}
        \item If $\mathsf{VerifyOfflineCompletion}(\mathsf{hist}_{\mathsf{rel}}, \mathsf{scm}^{\mathsf{rec}}_{k-1})$ and $\mathsf{VerifyOfflineCreation}(\mathsf{hist}_{\mathsf{rel}}, \mathsf{ccm}_{k})$ both return true, return true.
        \item Else, return false.
    \end{enumerate}
\end{func}
\begin{func}{$\mathsf{VerifyOfflineCreation}(\mathsf{hist}_{\mathsf{rel}}, \mathsf{scm}^{\mathsf{sen}}_{l})$}
    \item Let $m_{l}^{\mathsf{hist}}$ be the element with state commitment $\mathsf{scm}^{\mathsf{sen}}_{l}$ in $\mathsf{hist}_{\mathsf{rel}}$
    \item If $\mathsf{VerifyState}(m_{l}^{\mathsf{state}})$ is true, return true. %
    \item If $m_{l}^{\mathsf{hist}} = (\mathsf{scm}^{\mathsf{sen}}_{l}, \sigma^{\mathsf{sen}}_{l})$ and $\sigma^{\mathsf{sen}}_{l}$ is a valid central bank signatures for $\mathsf{scm}^{\mathsf{sen}}_{l}$, return true. %
    \item Parse $(\mathsf{sn}_{l}, \mathsf{ds}_{l}, \mathsf{scm}^{\mathsf{sen}}_{l}, \mathsf{dcm}^{\mathsf{sen}}_{l}, \mathsf{zkp}^{\mathsf{state}})$ from $m_{l}^{\mathsf{hist}}$
    \item Verify $\mathsf{zkp}^{\mathsf{state}}$ for public input $\mathsf{sn}_{l}, \mathsf{ds}_{l}, \mathsf{scm}^{\mathsf{sen}}_{l}, \mathsf{dcm}^{\mathsf{sen}}_{l}$
    \item Parse $(\mathsf{blind}_{l}^{\mathsf{dep}}, \mathsf{scm}^{\mathsf{sen}}_{l-1})$ from $m_{l}^{\mathsf{hist}}$%
    \item Verify that $\mathsf{dcm}^{\mathsf{sen}}_{l}$ equals to $\mathsf{comm}_{\mathsf{blind}_{l}^{\mathsf{dep}}}(\mathsf{scm}^{\mathsf{sen}}_{l-1})$
    \item If $\mathsf{scm}^{\mathsf{sen}}_{l-1}$ belongs to a payment creation state, return $\mathsf{VerifyOfflineCreation}(\mathsf{hist}_{\mathsf{rel}}, \mathsf{scm}^{\mathsf{sen}}_{l-1})$. \\
    Else return $\mathsf{VerifyOfflineCompletion}(\mathsf{hist}_{\mathsf{rel}}, \mathsf{scm}^{\mathsf{sen}}_{l-1})$ 
\end{func}
  \end{multicols}
  \end{mdframed}
\end{minipage}
    \caption{Pseudocode for state verification functions.}
    \label{fig:pseudo_ver}
\end{figure*}

\end{document}